\documentclass{JHEP3}

\input{epsf}
\usepackage{epsfig}
\usepackage{epstopdf}
\usepackage{amssymb}
\usepackage{amsfonts}
\usepackage{amsbsy}
\usepackage{graphicx}

\def\Ab{{\bf A}}
\def\a{{\bf A}}
\def\Bb{{\bf B}}
\def\b{{\bf B}}
\def\Cb{{\bf C}}
\def\c{{\bf C}}

\def\IC{\mathbb{C}}

\def\Z{\mathbb{Z}}
\def\iz{\mathbb{Z}}

\newcommand{\junk}[1]{}

\newcommand{\orbifold}{T^6/\Z_2 \times\Z_2}

\providecommand{\Ov}[1]{\frac{1}{#1}}

\renewcommand{\Re}{{\rm Re }}

\title{Diversity in the Tail\\ of the 
Intersecting Brane Landscape}

\author{Vladimir Rosenhaus and Washington Taylor\\
Center for Theoretical Physics\\
MIT\\
Cambridge, MA 02139, USA\\
{\tt vladr, wati} {\rm at} {\tt mit.edu}
}

\abstract{Techniques are developed for exploring the complete space of
  intersecting brane models on an orientifold.  The classification of
  all solutions for the widely-studied $T^6/{\Z}_2 \times {\Z}_2$
  orientifold is made possible by computing all combinations of branes
  with negative tadpole contributions.  This provides the necessary
  information to systematically and efficiently identify all models in
  this class with specific characteristics.  In particular, all ways
  in which a desired group $G$ can be realized by a system of
  intersecting branes can be enumerated in polynomial time.  We
  identify all distinct brane realizations of the gauge groups $SU(3)
  \times SU(2)$ and $SU(3) \times SU(2) \times U(1)$ which can be
  embedded in any model which is compatible with the tadpole and SUSY
  constraints.  We compute the distribution of the number of
  generations of ``quarks'' and find that 3 is neither suppressed nor
  particularly enhanced compared to other odd generation numbers.  The
  overall distribution of models is found to have a long tail.
  Despite disproportionate suppression of models in the tail by
  K-theory constraints, the tail of the distribution contains much of
  the diversity of low-energy physics structure.}

\preprint{MIT-CTP-4033}

\begin{document}

\section{Introduction}

Intersecting Brane Models \cite{IBM} have been studied for many years
as a simple class of string theory constructions giving rise to
low-energy physics theories containing a number of desirable features.
In particular, these models give rise to low-energy 4-dimensional
gauge theories with chiral fermions \cite{Bachas, bdl}, and have been
shown to include supersymmetric models with quasi-realistic
phenomenology.  For reviews of the subject of intersecting brane
models, see \cite{Uranga, Kiritsis, Lust, bcls, bkls, Marchesano, Lust-review}.
Recently, much work has been done on understanding nonperturbative
instanton effects in these models, which are relevant for computing
Yukawa couplings and for supersymmetry breaking.  These developments
are reviewed in \cite{bckw}

In this paper we carry out a detailed analysis of a well-studied class
of string compactifications, namely intersecting brane models on the
toroidal $T^6/\Z_2 \times \Z_2$ orientifold \cite{t6o}.  These models
are computationally very simple, and have been explored extensively.
Supersymmetric models have been found in this class with 3 generations
of matter in a standard model-like structure \cite{csu1, csu2, cim,
cl3, Marchesano-Shiu, cll}.

The mathematical structure of the vacuum classification problem for
intersecting brane models is similar in many ways to that of other
string vacuum constructions, such as flux compactifications in
type IIB string theory.  In all these cases, supersymmetry conditions
imply a positive-definite constraint on the topological degrees of
freedom encoding the vacuum configuration ({\it i.e.} brane windings
or fluxes).  Topological constraints give a limit to the total tadpole
contribution from these combinatorial degrees of freedom, so that the
mathematical problem of classifying vacua becomes one of solving a
partition problem.

For the intersecting brane models we consider here, this partition
problem is complicated by the fact that configurations are allowed in
which some tadpoles become negative.  These branes with negative
tadpoles make even a proof of finiteness of vacuum solutions rather
nontrivial.  Early work on IBM's on the $\orbifold$ orientifold
focused on solutions with certain physical properties.  Some
systematic analysis of models looking for solutions with standard
model-like properties was done in \cite{cll}.  A systematic computer
search through the space of all solutions was carried out in
\cite{Blumenhagen,Gmeiner}.  In some 45 years of computer time (using
clusters), 1.66 $\times 10^{8}$ consistent models were identified.
The results of this analysis suggested that the gauge group and number
of generations of various matter fields were essentially independently
and fairly broadly distributed, without constraints (up to some upper
bounds) in the space of available models, and estimates were given for
the frequency of occurrence of various physical features in the
models.  In \cite{Douglas-Taylor}, the role of branes with negative
tadpole contributions (``{\bf A}-branes'')\footnote{Such branes were
referred to as ``NZ'' branes in \cite{cll}, and ``type IV'' branes in
\cite{Cvetic:2002pj}; we will follow the notation of
\cite{Douglas-Taylor} here, and apologize for confusion due to the
variety of prior conventions in the literature.} was systematically
analyzed.  It was proven there that the total number of supersymmetric
IBM vacuum solutions in this model is finite.  Further evidence was
given for the broad and independent distributions of gauge group
components and numbers of matter fields.  Furthermore, analytic tools
and estimates were given for numbers of brane configurations with
certain properties.

In this paper we complete the program of analysis begun in
\cite{Douglas-Taylor}.  We systematically analyze the possible
configurations of negative-tadpole {\bf A}-branes which can arise in
models of this type.  We develop polynomial time algorithms for classifying
all {\bf A}-brane configurations, and numerically determine that there
are precisely 99,479 such configurations satisfying supersymmetry and
tadpole constraints.  For each of these {\bf
A}-brane configurations, the distribution of remaining branes is given
by a partition problem with all branes contributing positive amounts
to all tadpoles, so the complete range of models with any desired
properties can be carried out in a straightforward fashion given the
data on {\bf A}-brane combinations.    

The analysis in this paper is a prototype for other classes of models,
such as magnetized brane models on Calabi-Yau manifolds, which may
present similar challenges for analyzing vacua due to negative tadpole
contributions.  The results on \a-brane configurations enable us to
systematically analyze all IBM models on the orbifold of interest with
specific physical features.  In particular, it is possible to
efficiently classify all ways in which a particular gauge group $G$
can appear as a subgroup of the full gauge group.  We carry out this
analysis for the gauge subgroup $G =$ SU(3) $\times$ SU(2), and find
some 218,379 distinct ways in which this group $G$ can be realized as
a subgroup of the gauge group while satisfying SUSY and
undersaturating the tadpole constraints.
These constructions can be extended to some 16 million distinct
realizations of $SU(3) \times SU(2) \times U(1)$, which we have also
enumerated.  We look at the number of generations of ``quarks'' in the
({\bf 3, 2}) representation of the gauge group, and find that this
generation number generally ranges from 1 to about ${\cal O} (10)$, is
peaked at 1 and ranges out to 100 or so, with no particular
suppression or enhancement of 3 generations relative to other odd
generation numbers.  In principle all of  these constructions of gauge
groups of interest can be extended to all complete models containing
each realization.  K-theory constraints  then substantially reduce the
number of total possible models.

It is interesting to compare the results of our analysis with those of
Gmeiner, Blumenhagen, Honecker, L\"ust and Weigand \cite{Gmeiner}.
Those authors carried out a computer scan through models by looking at
configurations with small values of toroidal moduli (converted to
integers).  Their program generated a large class of models arising
from constructing all possible combinations of the set of branes
compatible with each fixed set of small moduli.  While there are many
more models for a typical set of small moduli than a typical set of
large moduli, we find that the full distribution of models has a very
long tail.  There are many configurations, particularly those with one
or more {\bf A}-branes, which have large values of the
(integer-converted) moduli.  For these larger moduli, the number of
combinatorial possibilities for models is smaller than for small
moduli.  Nonetheless, there are many distinct moduli for which some
models are possible, and the configurations of branes in the tail have
a wider range of variability.  For example, most of the SU(3) $\times$
SU(2) models we have found lie outside the range of moduli scanned in
\cite{Gmeiner}.  The long tail of the distribution and the increased
diversity of configurations in the tail explain how it is that the
computer search by Gmeiner {\it et al.} may have covered the region of
moduli space containing the greatest total number of models, and nonetheless
did not encounter over 95\% of the possible ways in which an SU(3)
$\times$ SU(2) $\times$ U(1) gauge subgroup can be constructed.  In
general, we expect the wider diversity of models in the tail to lead
to a greater probability of generating models with properties of
specific physical interest.  So, the answer to a question about
``generic'' properties of a typical model will depend crucially on how
the question is posed.  For example if we sample all IBM models which
saturate the tadpole conditions and ask for generic properties of
random models with given gauge subgroup $G$, we may get a very
different answer than if we sample all distinct ways in which the
gauge subgroup $G$ can be realized, independent of the number of other
components which can be included in the gauge group through extra
branes.  Our results suggest in particular that models with a desired
gauge group and specific other features (like a particular generation
number, no chiral exotic fields, etc.) may be more likely to be found
in the ``tail'' of the distribution than in the ``bulk''.

The structure of this paper is as follows.  In Section
\ref{sec:review} we review earlier work on intersecting brane models
on the toroidal orientifold of interest and define terminology and
notation needed for the analysis of the rest of the paper.  Section
\ref{sec:a} contains a complete analysis of the range of possible {\bf
A}-brane configurations.  In Section \ref{sec:fixed-gauge} we
demonstrate how all configurations realizing a desired gauge subgroup $G$ can be
enumerated in polynomial time, and give the results of such an
enumeration for $G =$ SU(3) $\times$ SU(2)
and $SU(3) \times SU(2) \times U(1)$.  Section
\ref{sec:comparison} contains a comparison of our results with those
of the moduli-based search of \cite{Gmeiner}, and a discussion of
the ``tail'' of the vacuum distribution.  We conclude in Section
\ref{sec:conclusions} with a summary and discussion of further related
questions.

\section{Review of Intersecting Brane Models}
\label{sec:review}

The general structure of intersecting brane models is reviewed in, for
example, \cite{bcls}.  We briefly review some of the basic structure
of these models relevant for this paper.  For the most part we follow
the notation and conventions of \cite{bcls, Douglas-Taylor}.

\subsection{General IBM's}
\label{sec:general-IBM}

Intersecting brane models are constructed by considering a string
compactification on some Calabi-Yau manifold, including branes which
can be wrapped around topologically nontrivial cycles in the manifold.
The situation of interest here involves supersymmetric configurations
of D6-branes wrapped on supersymmetric (special Lagrangian) three-cycles on the
Calabi-Yau.  Such brane configurations arise when type IIA string theory is
compactified on a Calabi-Yau with an orientifold six-plane (O6-plane);
the D6-branes are needed in this situation to cancel the negative
Ramond-Ramond charge carried by the O6-plane.  At points where the
branes intersect, there are massless string states giving chiral
fermions in the four-dimensional low-energy theory in the
uncompactified directions \cite{bdl}.

\subsection{$T^6/\iz_2 \times \iz_2$}
\label{sec:t6}

The specific class of models which we analyze in this paper arise from
an orientifold on a particular orbifold limit of a Calabi-Yau.
Situations where a singular limit of a Calabi-Yau can be described as
a toroidal orbifold are particularly easy to analyze and have long
been used as the simplest examples of string compactifications with
given amounts of supersymmetry.  In this case we consider the toroidal
orbifold $\orbifold$.  Considering the $T^6$ as a product of three
2-tori with complex coordinates $z_i, i = 1, 2, 3$, the orbifold
group is generated by the actions
\begin{equation}
\rho_1: (z_1, z_2, z_3) \rightarrow(-z_1, -z_2, z_3), \;\;\;\;\;
\rho_2: (z_1, z_2, z_3) \rightarrow(-z_1, z_2, -z_3).
\end{equation}
The product of these generators gives an additional element of the
orbifold group $\rho_3 = \rho_1 \rho_2:
(z_1, z_2, z_3) \rightarrow(z_1, -z_2, -z_3)$.
The geometric part of the orientifold action is given by
\begin{equation}
\Omega:z_i \rightarrow \bar{z}_i \,.
\end{equation}
The symmetry of the $T^6$ under the orbifold and orientifold actions
fixes many of the moduli parameterizing the shape of the torus.
Symmetry under the orbifold group guarantees that the torus factorizes
as a product of three 2-tori.  Symmetry under the orientifold action
constrains the complex structure of each torus.  For each torus the
complex structure $\tau = a + ib$ must map to $\bar{\tau} = a-ib$,
which must be in the 2D lattice generated by ($1, \tau$), so $\tau +
\bar{\tau} = 2a$ must be in the lattice.  This implies that either $a
= 0$ or $a = 1/2$, so that the torus is either rectangular or tilted
by one half cycle.
\vspace*{0.1in}

\noindent
{\bf Wrapped branes and tadpoles}

Consider first the case where $T^6$ is the product of three
rectangular 2-tori.  In this case supersymmetric D6-branes associated
with special Lagrangian 3-cycles on the $T^6$ which are invariant under the
orbifold group can be described in terms of winding numbers $(n_i,
m_i)$ on the three 2-tori.  A brane which cannot be decomposed into
multiple copies of a brane with smaller winding numbers has $(n_i,
m_i)$ relatively prime for all $i$, and is known as a {\it primitive}
brane.  Each brane has an image under the orientifold action with
winding numbers $(n_i, -m_i)$.  

There are Ramond-Ramond charges associated with the O6-plane which
must be cancelled by the wrapped D6-branes for consistency of the
model in the absence of other Ramond-Ramond sources (such as fluxes).
Each element $\rho$ in the orbifold group (including the identity)
gives rise to an orientifold transformation $\Omega_\rho = \Omega
\rho$ which, coupled with world-sheet orientifold reversal, gives a
symmetry of the string theory and is associated with an orientifold
charge on the fixed plane of $\Omega_\rho$.  This gives rise to 4
independent tadpole cancellation conditions, associated with total
6-brane charge along the directions $(x_1 x_2 x_3),  (x_1, y_2, y_3),
(y_1, x_2, y_3), (y_1, y_2, x_3)$.  Labeling these tadpoles $P, Q, R,
S$, each brane with winding numbers $(n_i, m_i)$ contributes to the
tadpoles (with appropriate sign conventions)
\begin{eqnarray}
P & = &  n_1 n_2 n_3 \nonumber\\
Q & = &  -n_1 m_2m_3 \nonumber\\
R & = &  -m_1 n_2m_3 \label{eq:tadpoles}\\
S & = &  -m_1 m_2n_3 \nonumber\,.
\end{eqnarray}
The cancellation of tadpoles requires that summing over all branes,
indexed by $a$, must give
\begin{equation}
\sum_{a}P_a =\sum_{a}Q_a= \sum_{a}R_a= \sum_{a}S_a=  T = 8
\label{eq:total-tadpole}
\end{equation}
where $-T = -8$ is the contribution to each tadpole from the O6-plane.
\vspace*{0.1in}
When giving explicit examples of branes we will generally indicate the
brane by the values of the tadpoles, with winding numbers as
subscripts, in the form
\begin{equation}
(P, Q, R, S)_{(n_1, m_1; n_2, m_2; n_3, m_3)} =
(-1, 1, 1, 1)_{(1, 1; 1, 1; -1, -1)}
\end{equation}

\noindent
{\bf Supersymmetry conditions}

Supersymmetry imposes further conditions on the winding numbers of the
branes.  The three moduli associated with the shapes of the three
2-tori can be encoded in three positive parameters\footnote{These parameters are $j = j_2 j_3, k = j_1 j_3, l = j_1
  j_2$ in terms of the (imaginary) toroidal moduli $\tau_k = i/j_k$
  for rectangular tori.}
 $j, k,l$, in terms
of which the supersymmetry conditions become
\begin{equation}
m_1 m_2m_3-jm_1 n_2n_3-kn_1 m_2n_3-ln_1 n_2m_3 = 0
\label{eq:SUSY-1}
\end{equation}
and
\begin{equation}
P + \frac{1}{j}  Q + \frac{1}{k}  R + \frac{1}{l}  S > 0 
\label{eq:SUSY-2}
\end{equation}
for each brane separately, with the same positive
values of the moduli $j, k, l$ for all 
branes. When all tadpoles are nonvanishing, (\ref{eq:SUSY-1}) can be rewritten as
\begin{equation}
\frac{1}{P}  + \frac{j}{Q}  + \frac{k}{ R}  + \frac{l}{S}  = 0  \,.
\label{eq:SUSY-1b}
\end{equation}
Finally, there is a further discrete constraint from K-theory which
states that when we sum over all branes we must have
\cite{Uranga:2000xp}
\begin{equation}
\sum_{a} m^a_1 m^a_2 m^a_3 \equiv
\sum_{a} m^a_1  n^a_2  n^a_3 \equiv
\sum_{a}  n^a_1 m^a_2  n^a_3 \equiv
\sum_{a} n^a_1 n^a_2 m^a_3 \equiv
0 \; ({\rm mod} \; 2) \,,
\label{eq:K-theory}
\end{equation}
where $n^a_i, m^a_i$ give the  winding numbers of the $a$th brane on
the $i$th torus.
As we will see, this discrete constraint significantly decreases the
extreme end of the tail of the distribution of allowed models on the
moduli space.
\vspace*{0.1in}

\noindent
{\bf Branes on tilted tori}

Now, we return to the case where the torus is tilted, with $\Re \;\tau =
1/2$.
Say the $i$th torus is tilted.
In this case, we can define winding numbers $\hat{n}_i, \hat{m}_i$
around generating cycles $[a_i], [b_i]$ of the tilted torus
\footnote{Following the conventions of \cite{bcls}, these generating
  cycles are given on the complex plane by $2 \pi (R^i_1 + i R^i_2/2),
  2 \pi i R^i_2$.}.  In terms of
these winding numbers, we can define
\begin{equation}
n_i = \hat{n}_i, \;\;\;\;\; \tilde{m}_i = \hat{m}_i +\frac{1}{2} \hat{n}_i
\label{eq:orthogonal-nm}
\end{equation}
which represent the number of times the brane winds along the
perpendicular $x, y$ axes on the $i$th torus.  
The supersymmetry conditions for a brane on a tilted
  torus are again (\ref{eq:SUSY-1}, \ref{eq:SUSY-2}) in terms of $n_i,
  \tilde{m}_i$ defined in (\ref{eq:orthogonal-nm}).  The tadpole
  conditions on a tilted torus are given by (\ref{eq:total-tadpole})
  where we use $m_i = 2 \tilde{m}_i$ on the tilted tori
  \cite{Cvetic:2002pj}.  A difference which arises on the tilted
  torus is that the range of values allowed for $(n_i, \tilde{m}_i)$
  is different from the condition of relatively prime integers imposed
  on the winding numbers for the rectangular torus.  On the tilted
  torus, the winding numbers $\hat{n}_i, \hat{m}_i$ must be relatively
  prime integers.  Integrality of $\hat{n}_i, \hat{m}_i$ imposes the
  constraint that on a tilted torus we must have $n_i \equiv 2
  \tilde{m}_i ({\rm mod} \; 2)$.  The relative primality constraint on
  $\hat{n}_i, \hat{m}_i$ becomes the condition that $n_i,
  2\tilde{m}_i$ have no common prime factor $p > 2$, while for $p = 2$
  we can have $n_i$ and $2\tilde{m}_i$ both even iff $n_i/2$ and
  $\tilde{m}_i$ are not congruent mod 2.  Because of the common form
  of the tadpole relations, enumeration of branes on tilted tori is
  closely related to that of branes on rectangular tori, with some
  minor modifications from the modified relative primality constraint.
\vspace*{0.1in}

\noindent
{\bf Model construction}

We can now summarize the degrees of freedom and necessary conditions
on these degrees of freedom which must be satisfied to construct an
intersecting brane model on $\orbifold$.  First, we can have anywhere
from 0-3 tilted tori, with the remaining tori rectangular.  Then, we
wrap any number of branes on the tori, described by winding numbers
$n_i, m_i, i = 1, 2, 3$ for each brane, subject to the appropriate
primitivity conditions for rectangular/tilted tori, so that the total
tadpole from the branes is (\ref{eq:total-tadpole}).  The K-theory
constraints (\ref{eq:K-theory}) must be satisfied by the total brane
configuration.  Finally, moduli $j, k, l$ must be chosen so that the
supersymmetry conditions (\ref{eq:SUSY-1}) and (\ref{eq:SUSY-2}) are
satisfied for each brane.  Note that in general 3 branes will be
sufficient to completely constrain the moduli, which are then rational
numbers since the constraint equations are linear with rational
coefficients.  In some cases with fewer branes or redundant constraint
equations, there are one or two remaining unfixed moduli.  In these
cases we can always choose representative combinations of moduli which
are rational, though this choice is not unique.
\vspace*{0.1in}

\noindent
{\bf Symmetries}

The $\orbifold$ model has a number of symmetries under which related
models should be identified.  Permutations of the three tori can give
arbitrary permutations on the indices $i = 1, 2, 3$ of the winding
pairs $(n_i, m_i)$, and hence the same permutation on the tadpoles $Q,
R, S$ and moduli $j, k, l$.  By 90 degree rotations $n_i \rightarrow
m_i \rightarrow -n_i$ on two of the tori, we have a further symmetry
under exchange of $P$ with any of the other tadpoles.  This extends
the symmetry to the full permutation group on the set of 4 tadpoles.
(To realize this symmetry on the moduli it is convenient to write the
moduli as $j/h, k/h, l/h$, so that $h$ plays a symmetric role to the
other moduli.  When the original moduli are rational, we can then
uniquely choose $h, j, k, l$ to be integers without a common
denominator.  We will go back and forth freely between these two
descriptions in terms of 3 rational or 4 integral moduli.)

There is a further set of symmetries on the winding numbers which do
not affect the tadpoles.  We can rotate two tori by 180 degrees,
changing sign on $n_i, m_i$ for two of the $i$'s.  Each brane also has
an orientifold image given by negating all $m_i$'s.  We will keep only
one orientifold copy of each brane, and fix the winding number
symmetries as in section 2.2 of \cite{Douglas-Taylor} by keeping only
branes with certain combinations of winding number signs, as described
there.

\vspace*{0.1in}

\noindent
{\bf Types of branes}

There are three distinct types of branes which are compatible with the
supersymmetry conditions.  These types are distinguished by the
numbers of nonzero tadpoles and the signs of the tadpoles.  
The
allowed brane types are:
\vspace*{0.06in}

\noindent
{\bf A}-branes: These branes have 4 nonzero tadpoles, of which one is
negative and 3 positive.  An example of an \a-brane is
\begin{equation}
(P, Q, R, S)_{(n_1, m_1; n_2, m_2; n_3, m_3)} =
(-1, 1, 2, 2)_{(1, 2; 1, 1; -1, -1)} \,.
\end{equation}

\noindent
{\bf B}-branes: These branes have 2 nonzero tadpoles, both positive.
An example of a \b-brane is
\begin{equation}
(P, Q, R, S)_{(n_1, m_1; n_2, m_2; n_3, m_3)} =
(1, 0, 0,
1)_{(1, 1; 1, -1; 1, 0)}  \,.
\end{equation}

\noindent
{\bf C}-branes: These branes have only 1 nonzero tadpole, and are
wrapped on cycles associated with the O6 charge.
An example of a \c-brane is
\begin{equation}
(P, Q, R, S)_{(n_1, m_1; n_2, m_2; n_3, m_3)} =
(1, 0, 0,0)_{(1, 0; 1, 0; 1, 0)} \,.
\end{equation}
\c-branes are often referred to as ``filler'' branes in the
literature, since they automatically satisfy the SUSY conditions and
can be added to any configuration which undersaturates the tadpoles to
fill up the total tadpole constraint.
\vspace*{0.1in}

\noindent
{\bf Low-energy gauge groups and matter content}

Given a set of branes and moduli satisfying the tadpole,
supersymmetry, and K-theory constraints, the gauge group and matter
content of the low-energy 4-dimensional field theory arising from the
associated compactification of string theory can be determined from
the topological structure of the branes.  

A set of $N$ identical branes of type \a\ or \b\ give rise to a U(N) gauge
group in the low-energy theory, from the $N \times N$ strings
stretching between the branes.  A set of $N$ identical type \c\ branes,
on the other hand, which are coincident with their orientifold images,
give rise to a group\footnote{There are a variety of notations for
symplectic groups in the math and physics literature.  By Sp(N) we
denote the symplectic group of $2 N \times 2 N$ matrices composing the
real compact Lie group whose algebra has Cartan classification $C_N$,
which can be defined as $Sp (N) =U(2N) \cap Sp (2 N,\IC)$.  Note that
this group is referred to by some authors as $USp (2 N)$, and by some
authors (such as in \cite{bcls}) as $Sp (2 N)$.}
Sp(N).

Associated with each pair of branes there are matter fields containing
chiral fermions associated with strings stretching between the branes.
These matter fields transform in the bifundamental of the two gauge
groups of the branes.  The number of copies (generations) of these
comes from the intersection number between the branes.  For two branes
with winding numbers $n_i, m_i$ and $\check{n}_i, \check{m}_i$, the
intersection number is
\begin{equation}
I = \prod_{i}(n_i {\check{m}}_i-{\check{n}}_im_i) \,.
\label{eq:intersection}
\end{equation}
Because of the orientifold, there is a distinction between \Ab\ and
\Bb\ type branes $a, b$, which have images $a' \neq a$ and $b' \neq b$
under the action of $\Omega$, and a \c\ type brane $c$, which is taken
to itself under the action of $\Omega$.  Given two \Ab-type branes $a,
\hat{a}$, for example, the intersection numbers $I_{a \hat{a}}$ and
$I_{a\hat{a}'}$ are distinct, and must be computed separately,
corresponding to matter fields in the fundamental and antifundamental
representations of the gauge group on the branes $\hat{a}$.  The same
is true of type \b\ branes, but not type \c\ branes, which are equal to
their orientifold images.  Note that on tilted tori, in the
intersection formula (\ref{eq:intersection}), the winding number
$\tilde{m}_i$ is used in place of $m_i$.  Note also that for branes
$a, d$ of any type the parity of $I_{ad}$ and $I_{ad'}$ are the same,
so that the sum $I_{ad} + I_{ad'}$ is always even.  This means that if
there is a stack of $N$ branes $a$ and 2 branes $d$ of type \a\ or \b,
the number of matter fields in the fundamental of $SU(N)$ and the
fundamental of $SU(2)$ (which is equivalent to the antifundamental) is
even unless there is a tilted torus
in at least one dimension.

\subsection{The space of SUSY IBM models}
\label{sec:space-models}

As reviewed in \cite{bcls}, the first intersecting brane models with
chiral matter which were constructed lacked supersymmetry.  There are
an infinite number of such models, which generally have perturbative
or nonperturbative instabilities.  In this paper we restrict attention
to supersymmetric models, which have a more robust structure.

For the IBM models on the $\orbifold$ orientifold described above, the
problem of constructing supersymmetric models amounts to solving a
partition problem.  If we have a set of branes indexed by $a$, for
each $a$ the tadpoles form a 4-vector of integers $(P_a, Q_a, R_a,
S_a)$.  The tadpole constraint  says that the sum of these vectors
must equal the vector $(T, T, T, T) = (8, 8, 8, 8)$ associated with
the orientifold charges.  If all tadpoles $P_a, \cdots, S_a$ were
nonnegative, with each brane having at least one positive tadpole,
then the number of solutions of the associated partition problem would
obviously be finite, as the number of possible tadpole charges would
be at most $8^4 -1 = 4095$, a subset of which would be described by a
nonzero but finite number of winding number configurations, and the
maximum total number of branes would be 32.  With some branes allowed
to have one negative tadpole, however, it is no longer so clear that
the number of solutions of the partition problem must be finite, or
even that the number of branes in any solution is bounded.

For fixed values of the moduli $j, k, l$, it is fairly straightforward
to demonstrate that the number of solutions is finite \cite{Blumenhagen}.  The SUSY
inequality (\ref{eq:SUSY-2}) states that the linear combination of
tadpoles $\gamma_a =P_a + Q_a/j + R_a/k + S_a/l$ is positive for every
brane.  The total of this quantity over all branes must be $T (1
+ 1/j + 1/k + 1/l)$, and is therefore bounded for fixed moduli.
The contribution to $\gamma_a$ from each brane is bigger than any of
the individual contributions from any positive term.  To see this,
assume for example that  $Q_a > 0$, and we will show that $\gamma_a >
Q_a/j$.  Assume without loss of generality that $P_a = -P< 0$.
From the SUSY equality
(\ref{eq:SUSY-1b}) we have $k/R_a < 1/P$, so $R_a/k > P$.  Recalling that
at most one tadpole is negative, we then have $ \gamma_a > Q_a/j +
R_a/k-P> Q_a/j$.  Thus, for fixed moduli, we have bounded each
individual positive tadpole contribution.  But since each winding
number appears in two tadpoles, this bounds all winding numbers.  It
follows that there are a finite number of different possible branes
consistent with the SUSY conditions for fixed moduli $j, k, l$.  There
is therefore a minimum required contribution to $\gamma_a$, and
therefore a maximum number of branes which can be combined in any
model saturating the tadpole conditions, proving that there are a finite
number of models for fixed moduli.

In \cite{Blumenhagen, Gmeiner}, the space of solutions was scanned by
systematically running through moduli and finding all solutions
saturating the tadpole conditions and solving the SUSY and K-theory
constraints for each combination of moduli.  Writing the moduli as a
4-tuple of integers $\vec{U} = (h, j, k, l)$ as discussed above, they
scanned all solutions up to $| \vec{U} | = 12$. 
This computer
analysis produced some 1.66 $\times 10^{8}$ SUSY solutions.
Their
numerical results indicated that the number of models was decreasing
fairly quickly as the norm of the moduli vector increased, so that
this set of models seemed to represent the bulk of the solution space.

In \cite{Douglas-Taylor}, an analytic approach was taken to analyzing
the space of SUSY models.  In this paper it was demonstrated using the
SUSY conditions that even including all possible moduli, the total
number of brane configurations giving SUSY models saturating the
tadpole condition is finite.  Estimates were found for numbers of
models with particular brane structure.

In this paper we complete this analysis.  The key to constructing all
models with some desired structure is to deal with the {\bf A}-branes
systematically.  In the next section, we describe how all 99,479
distinct {\bf A}-brane combinations which do not over-saturate the
tadpole conditions can be constructed.  For each of these
combinations it is then straightforward, if tedious, to construct all
models which complete the tadpole conditions through addition of type
{\bf B} and {\bf C} branes.

\section{Constructing {\bf A}-brane configurations}
\label{sec:a}

As discussed above, finding all models in which a combination of
branes satisfy the SUSY and tadpole conditions for some set of moduli
would be straightforward if all branes had only positive tadpoles.  In
this situation, all tadpoles in each brane would give positive
contributions to the tadpole condition.  This would give us upper
bounds on the winding numbers for all the branes and on the maximum
number of branes in a given configuration.  This would then allow us to
scan over all allowed winding numbers and hence find all possible
models, or all models with some particular desired properties.

Since \Ab-branes have a negative tadpole, however, the tadpole
constraint alone will not give us upper bounds on the winding numbers.
In order to obtain such upper bounds we need to use a combination of
the tadpole condition and the SUSY condition.  Indeed, this approach
was used in \cite{Douglas-Taylor} to prove that the total number of
supersymmetric brane configurations satisfying the tadpole condition
is finite.  The bounds determined in that paper, however, are too
coarse to allow a search for all models in any reasonable amount of
time.  In this section we obtain tighter bounds, and describe how
these can be used to implement a systematic search for all allowed
\Ab-brane combinations.  We have carried out such a search, and
describe the results here.

The goal of this section is thus the construction of all possible
configurations of \Ab-branes compatible with the tadpole and SUSY
constraints.  Having all \Ab-brane configurations is a crucial step in
performing a complete search for any class of models.  In 3.1 we
develop analytic bounds on the various combinations of \a-branes with
different negative tadpoles.  These bounds are derived using
combinations of the tadpole and SUSY conditions in order to place
upper bounds on the winding numbers and the maximum number of branes
allowed in a configuration.  Then in 3.2 we apply these bounds to
construct a complete algorithm for generating all \Ab-brane configurations.  In
\ref{sec:a-results} we summarize the results of an exhaustive
numerical analysis of all the \a-brane combinations.  In this section
we will be working exclusively with \Ab-type branes.
In the following section we describe how to systematically add \b- and
\c-branes to form all configurations with desired physical properties.

To simplify the discussion we define some notation.  We let
$[p,q,r,s]$ denote a configuration consisting of $p$ branes with a
negative $P$ tadpole, $q$ branes with a negative $Q$ tadpole, $r$
branes with a negative $R$ tadpole and $s$ branes with a negative $S$
tadpole.  Through permutations in the ordering, we can always arrange
the branes so that the branes with a negative $P$ tadpole are first in
the configuration, followed by those with a negative $Q$ tadpole, then
$R$, then $S$.  Within these groupings the negative tadpoles can be
canonically ordered in increasing order of magnitude.  A subscript $a$ on a tadpole
($P_a$ etc.)  will indicate which brane it belongs to.  Also, for
convenience, in this section we will write all tadpole numbers as the
absolute value of the tadpole contribution, and explicitly insert the
minus sign when needed.  For instance the tadpoles of a brane $a \leq
p$ with a negative $P$ tadpole will be written as
$(-P_a,Q_a,R_a,S_a)$.  In the same manner, whenever we write a winding
number $n_i$ or $m_i$, it will mean the absolute value of the winding
number and we will explicitly insert a minus sign when needed.
Throughout this section we will work with four integer moduli $h, j,
k, l$ for symmetry in the equations.  Also, as in Section 2, the
tadpole bound of 8 is denoted by $T$.

\subsection{Bounds from SUSY and Tadpole Constraints}
\label{sec:bounds}

We will classify the \Ab-brane configurations in terms of how many
different types of tadpoles are negative.  There will be four cases to
consider: $[p,0,0,0]$ which corresponds to only the $P$ tadpole being
negative, $[p,q,0,0]$ in which the first $p$ branes have a negative
$P$ tadpole and the next $q$ branes have a negative $Q$ tadpole,
$[p,q,r,0]$, and $[p,q,r,s]$.

We begin with the simplest case, $[p, 0, 0, 0]$, for which we can
immediately derive bounds on the winding numbers.  We then
derive several general conditions which are useful in proving tight
bounds on the winding numbers in the cases $[p, q, 0, 0]$ and $[p, q,
  r, 0]$.  We end this subsection by finding strong bounds for the
winding numbers in the case $[p, q, r, s]$.  In fact, it turns out
that there are no combinations of \a-branes which include branes with
each of the four tadpoles being negative, so there are in fact no
allowed combinations of type $[p, q, r, s]$ with $p, q, r, s > 0$.
The details of the proof of this statement are given in the Appendix.
This result, however, makes the analysis of all \a-brane combinations
much easier, since it immediately indicates that there are at most 8
\a-branes in any combination (since all have positive tadpoles $S_a >
0$, with $\sum_{i}S_a \leq 8$).

As the simplest case, we now consider the $[p,0,0,0]$ combinations,
which contain $p$ branes with tadpoles $(-P_a, Q_a, R_a, S_a), 1 \leq
a \leq p$.  Since the Q, R, and S tadpoles are each at least 1, and
the sums of the Q, R, and S tadpoles must each be less than or equal
to $T$, there can be a maximum of $T$ branes per configuration.  More
explicitly, each brane has 6 winding numbers $n_i, m_i$ from which the
4 tadpoles are constructed as cubic combinations through
(\ref{eq:tadpoles}).  Each winding number appears in 2 different
tadpoles (for example, $n_1$ appears in the $P$ and $Q$ tadpoles).
All 6 winding numbers appear in at least one of the Q, R, or S
tadpoles.  Since all the negative tadpoles are $P$, all 6 of the
winding numbers for each brane can be bounded from the tadpole
constraint applied to the Q, R, and S tadpoles.  For instance, the
tadpole constraint tells us that the sum of the Q tadpoles is less
than or equal to T.  Thus, the sum over branes of $n_1 m_2m_3$ is
$\leq T = 8$, which gives a strong upper bound on the winding numbers
$n^a_1, m^a_2, m^a_3$.  We can easily enumerate the allowed $[p, 0, 0,
0]$ combinations by considering all winding numbers for the first
brane which give $Q_1, R_1, S_1 \leq T$, then constructing all second
branes which give $Q_1 + Q_2 \leq T, \ldots$, and so on up to at most
$T = 8$ branes.

The other cases, with more than one type of negative tadpole, require
a somewhat more complicated analysis.  In the case $[p,0,0,0]$ we did
not need to use the SUSY conditions.  In the cases where there is more
than one different negative tadpole, however,  it is not immediately clear
how to bound the number of branes in a configuration.  The SUSY
condition will play an essential role in this constraint.  Rather than
immediately analyzing the next case, where both $P$ and $Q$ tadpoles
can be negative, we will find it useful to
first derive several conditions which will be helpful to us for
bounding the winding numbers and number of branes in a general
configuration.

In the first condition that we derive we consider the subset
of branes in the configuration with negative $P$ or $Q$ tadpoles ({\it
i.e.}, the first $p + q$ branes).  Among the first $p + q$ branes in
any \a-brane configuration (ordered as described above) there are no
negative R or S tadpoles.  When considering constructions of the type $[p,q,0,0]$ we can therefore use the tadpole
constraints for the $R$ and $S$ tadpoles to bound 5 of the 6 winding
numbers.  $n_1$  is the only winding number
which is not immediately bounded by the tadpole conditions, since it
is the only one that does not appear in the $R$ or $S$ tadpoles.  
We now determine a bound for these
$n_1$ winding numbers.
\vspace*{0.1in}

\noindent
\textbf{First Winding Number Bound (FWNB)} 
\vspace*{0.1in}

Again, we consider the first $p + q$ branes, which are those having
only negative $P$ or $Q$ tadpoles.  We choose $a, b$ from $a \in
\left\{1,...,p\right\}$, $b \in \left\{p+1,...p+q\right\}$.  The SUSY
condition (\ref{eq:SUSY-1b}) for brane $a$ states
$-\frac{h}{P_a}+\frac{j}{Q_a} + \frac{k}{R_a} + \frac{l}{S_a}=0$.  We
must therefore have $-\frac{h}{P_a}+\frac{j}{Q_a} <0$.  Similarly,
from the SUSY condition for brane $b$ we get $\frac{h}{P_b}
-\frac{j}{Q_b} <0$.  These two conditions together give
$\frac{Q_a}{P_a} >\frac{Q_b}{P_b} \ \ \forall a, b$.  Let $\lambda =
\max_b\left\{\frac{Q_b}{P_b}\right\}$.  Then $\frac{Q_a}{P_a} >
\lambda \ \ \forall a$ and $\lambda \geq \frac{Q_b}{P_b} \ \ \forall
b$.  Note that $\lambda = m^b_2m^b_3/n^b_2n^b_3$ for some $b$, so
$\lambda$ is independent of the winding numbers $n_1$ for any brane.
From the tadpole conditions for $P$ and $Q$ (upon rearranging) we get:
\begin{equation}\label{eq:FWNB}
(Q_1-\lambda P_1) + ...  + (Q_p - \lambda P_p) +(\lambda P_{p+1} -Q_{p+1})+ ...  + (\lambda P_{p+q} - Q_{p+q}) \leq T(\lambda + 1)
\end{equation}

Since each term in parenthesis is nonnegative, and the terms from the
first $p$ branes are positive, we have a bound on the $n_1$ winding
number of each of the branes with negative $p$, given all the
remaining winding numbers $n_2, n_3, m_i$ for each of these branes.
By choosing $\lambda =\min_a \left\{ \frac{Q_a}{P_a} \right\}$, we can
similarly determine a bound on the $n_1$ winding numbers for the
branes with negative $Q_b$.  We will henceforth refer to these bounds
on $n_1$ (\ref{eq:FWNB}) as FWNB for convenience.
\vspace*{0.1in}

The bound (\ref{eq:FWNB}) makes the construction of all configurations
of type $[p, q, 0, 0]$ a straightforward exercise, similar to that
described above for combinations of type $[p, 0, 0, 0]$, as we
describe in more detail in the following subsection.  We can derive
another condition which will be useful for cases with 3 or 4 different
negative tadpoles.  Once again, we look at the subset of branes in
which only two types of tadpole are negative, without loss of
generality taking these to be the first $p + q$ branes in the
configuration where $-P_a, -Q_b$ are the negative tadpoles.
\vspace*{0.1in}

 \noindent
\textbf{Two Column SUSY Bound (TCSB)}
\vspace*{0.1in}

For any combination of branes,
the tadpole conditions for the R and S branes gives 
\begin{equation}
\sum{\frac{R}{k}} + \sum{\frac{S}{l}} \leq T (\Ov{k}+\Ov{l}),
\end{equation}

where when we do not use indices the sum is taken over all branes and the tadpoles (here R and S) have their correct negative or positive values. For a brane of the type $(P_a,Q_a,-R_a,S_a)$ SUSY gives
$\frac{h}{P_a}+\frac{j}{Q_a}-\frac{k}{R_a} +\frac{l}{S_a} =0$.  Hence
$-\frac{k}{R_a}+\frac{l}{S_a} < 0 $, or rearranging,
\begin{equation}
-\frac{R_a}{k}+\frac{S_a}{l} >0\,.
\label{eq:basic-relation}
\end{equation}
Similarly for branes of the form
$(P_a,Q_a,R_a,-S_a)$ there is a positive contribution
$\frac{R_a}{k}-\frac{S_a}{l} >0$.  

Thus, we have that, even when restricting to just the first $p + q$
branes where both $R$ and $S$ are positive,
\begin{equation} \label{sump}
\sum_{a = 1}^{p + q}{\frac{R_a}{k}} + \sum_{a = 1}^{p + q}{\frac{S_a}{l}} \leq
T (\Ov{k}+\Ov{l})
\end{equation}

The relation (\ref{sump}) is possible only if
\begin{equation}\label{eq:TCSB}
\sum_{a = 1}^{p + q}{R_a} \leq T \textrm{ or } \sum_{a = 1}^{p + q}{S_a} \leq T
\end{equation}
where the inequality is a strict $<$ if there is at least one brane
with a negative $R$ or $S$ tadpole.

The Two Column SUSY Bound (TCSB) is (\ref{eq:TCSB}).  This result
clearly generalizes to considering the subset of branes with any two
tadpoles being purely positive, in which case one of the two positive
sums must be $\leq T$.
As a consequence of the condition (\ref{eq:TCSB}), we gain useful
information about configurations with 3 types of negative tadpoles
($[p, q, r, 0]$).  
In particular, we can show that

\begin{equation} 
\textrm{\textit{The sum of the positive contributions to one of the
first three tadpoles is less than 3T}}
\label{eq:pqr2}
\end{equation}

To see this, without loss of generality, we let $h= \min \left\{h,j,k\right\}$.  We have that 
\begin{equation}
\sum{\frac{P}{h}}+\sum{\frac{Q}{j}} + \sum{\frac{R}{k}} \leq T(\Ov{h} + \Ov{j} +\Ov{k})
\end{equation}

For a brane of the form $(-P_a,Q_a,R_a,S_a)$ there will be a positive
contribution to the above sum since
$-\frac{P_a}{h}+\frac{Q_a}{j} >0$, which can be shown in a similar
fashion to (\ref{eq:basic-relation}) above.  From a brane with a tadpole other
than $P_a$ that is negative, say $(P_a,-Q_a,R_a,S_a)$, we have
$-\frac{Q_a}{j}+\frac{R_a}{k} >0$, and so $\frac{P_a}{h}$ will
contribute less than the total for that brane to the above sum.  Hence
we get
\begin{displaymath}
\sum_{a = p + 1}^{p + q + r + s}{\frac{P_a}{h}} < T
(\Ov{h}+\Ov{j}+\Ov{k}) \leq \frac{3T}{h} \,.
\end{displaymath}
This proves (\ref{eq:pqr2})

Along with the TCSB, this condition suffices to give strong bounds on
the winding numbers and the number of branes in a $[p,q,r, 0]$
configuration.  We describe the details of how these bounds are used
to determine the range of $[p, q, r, 0]$ configurations in the next
subsection.  Finally, we need to consider the case of $[p,q,r,s]$.  It
turns out that there are no configurations of this type which are
compatible with SUSY and the tadpole conditions with $T = 8$ (though
such configurations are possible at larger values of $T$).  The full
proof that there are no configurations of this type is given in the
Appendix.  Here we just derive a bound on the maximum number of
allowed branes.

\indent Suppose we have $[p,q,r,s]$.  Without loss of generality, let  $s \leq r \leq q  \leq p$ and for convenience of notation, we let $a=p,\ b=p+q,\  c=p+q+r, \ d= p+q+r+s$.  The sum of the tadpole conditions for R and S gives
\begin{equation}
\sum{\frac{R}{k}} + \sum{\frac{S}{l}} \leq T( \Ov{k}+ \Ov{l})
\end{equation}

Looking at the left side of this equation, and using relations like
(\ref{eq:basic-relation}) we have
\begin{eqnarray*}
\frac{R_1+...+R_b}{k} + \frac{S_1+...+S_{b}}{l} + \underbrace{(-\frac{R_{b+1}}{k}+\frac{S_{b+1}}{l})}_{>0}+...+ \underbrace{(\frac{R_{d}}{k} - \frac{S_d}{l})}_{>0}
\end{eqnarray*}

This is greater than $(p+q)(\Ov{k} + \Ov{l})$, so if $(p+q) \geq T$ the tadpole condition is violated.  So we have found that 
\begin{equation}
\textrm{\textit{For }} (p+q) \geq T \textrm{\textit{ there are no
    configurations }} [p,q,r,s]
\label{eq:pq8}
\end{equation}
This condition severely limits the number of branes that can be in a
configuration.  In fact, by various manipulations of the SUSY
equations, we can show there are no configurations of the form
$[p,q,r,s]$ when $T = 8$ (see Appendix for details).

\subsection{Algorithm}
Using the constraints derived in the previous subsection we can
construct algorithms whose complexity scales polynomially in $T$ to
generate all \Ab-brane configurations.  We outline such algorithms for
the three cases $[p,0,0,0]$, $[p,q,0,0]$, and $[p,q,r,0]$.  In each
case, the algorithm consists of scanning over all the winding numbers
for all the branes in the configuration subject to given bounds.  The
bounds on the size of the winding numbers and the number of branes in
the configuration are obtained by using combinations of the conditions
in the previous section.  The main challenge in constructing such
algorithms is having explicit bounds for all the winding numbers.  So
the focus of this discussion is on explaining how all winding numbers
can be constrained using the results of the previous subsection.
\vspace*{0.05in}

As we saw in 3.1, for $[p,0,0,0]$ we can simply loop over all
six winding numbers for each of the branes, all of which are
constrained by the tadpole condition.  Considering all winding numbers
for the first brane, and then subtracting the resulting contributions
for $Q_1, R_1, S_1$ from the tadpole conditions while assuming that
$P_1 \geq P_a, a > 1$ gives stronger bounds for the winding numbers of
the second brane, and so on.
\vspace*{0.05in}

Generating all \a-brane configurations of the form
$[p,q,0,0]$ is only slightly more involved.  We
loop over 5 of the 6 winding numbers for each brane (since all but
$n_1$ are constrained by the tadpole condition).  We then loop over
$n_1$ for each brane after bounding it through use of FWNB.  For
example, let us look at the simplest case of $[1,1,0,0]$.  The tadpole
condition for the S tadpole gives
(recall that in this section we are using $n_i, m_i$ to denote
absolute values of winding numbers, with signs put in explicitly)
\begin{equation}
m_1^1 m_2^1 n_3^1 + m_1^2 m_2^2 n_3^2 \leq T,
\end{equation}
and the tadpole condition for the R tadpole gives
\begin{equation}
m_1^1 n_2^1 m_3^1 + m_1^2 n_2^2 m_3^2 \leq T.
\end{equation}
With these two conditions we have bounds on all winding numbers except
for $n_1^1$ and $n_1^2$.  We can therefore easily loop over all these
winding numbers.  Concretely, this means we have 10 nested loops.  In
the first loop, we loop over all $m_1^1 < T$, in the second loop we
loop over all $m_2^1$ such that $m_1^1 m_2^1 <T$, and so on.  Finally,
we have two additional nested loops for $n_1^1$ and $n_1^2$.  The
maximum values for these are obtained from FWNB.  For $n_1^1$ we have
the constraint
\begin{equation}
n_1^1 (m_2^1 m_3^1 - \lambda n_2^1 n_3^1) \leq T (\lambda +1),
\end{equation}
where $\lambda = \frac{m_2^2 n_3^2}{n_2^2 n_3^2}$.  And for $n_1^2$ we have
\begin{equation}
n_1^2 (\lambda n_2^2 n_3^2- m_2^2 m_3^2) \leq T (\lambda +1),
\end{equation}
where $\lambda = \frac{m_2^1 n_3^1}{n_2^1 n_3^1}$.
The general case of $[p,q,0,0]$ is done by a similar application of FWNB.
\vspace*{0.05in} 
 
\indent The case $[p,q,r,0]$ requires the most work.  The idea is
to first generate three branes with all different negative tadpoles.
For the purpose of this discussion we rearrange the brane ordering so
that these branes can be chosen to have tadpoles
\begin{equation}
(-P_1, Q_1, R_1, S_1), \;\;\;\;\;
(P_2, -Q_2, R_2, S_2), \;\;\;\;\;
(P_3, Q_3, -R_3, S_3) \,.
\label{eq:3-branes}
\end{equation}
Note that further branes may be added to the configuration with
negative $P, Q,$ or $R$ tadpoles.  However, for any configuration we
choose the first brane to be the brane in that configuration with a
negative $P$ tadpole which minimizes $Q_a/P_a$, and the second brane
to be the one with a negative $Q$ tadpole which maximizes $Q_b/P_b$.
Once we have chosen the 3 branes (\ref{eq:3-branes}) we solve for
the moduli, and then  add all further combinations of branes consistent with those
moduli and the tadpole constraints.  

The first step is generating the first three branes,
(\ref{eq:3-branes}).  In order to generate these we need constraints
on all the winding numbers for these three branes.  From the tadpole
constraint on the S tadpole, since the total configuration of which
(\ref{eq:3-branes}) is a subset has no branes with negative $S$, we
have bounds on the $m_1$, $m_2$ and $n_3$ winding numbers for each of
the branes.  To find constraints for the other winding numbers we are
not allowed to use the tadpole bound condition on the $P,Q,$ or $R$
tadpoles, since the additional branes we may add to complete the
configuration can give negative contributions to these tadpoles.  We
will thus rely on the conditions derived in Subsection
\ref{sec:bounds} in order to specify upper bounds on the winding
numbers.  

We begin by using (\ref{eq:pqr2}).  Without loss of
generality this condition allows us to assume that $R_1+R_2 < 3T$.
Referring back to FWNB,  we take $\lambda =
Q_2/P_2$ which in (\ref{eq:FWNB}) gives  $(Q_1 - \lambda P_1) < T(\lambda +1)$
to constrain $n_1^1$.  (Note that since brane 2 has the largest value
of $Q_b/P_b$ among all negative $Q$ branes, all contributions on the
LHS of (\ref{eq:FWNB}) are positive, even when further branes are
included in the configuration.)

Next, we take $\lambda = \frac{Q_1}{P_1}$ and in a similar fashion use
$(\lambda P_2 - Q_2) <T(\lambda +1)$ to constrain $n_1^2$.  At this
point, the first two branes are completely constrained (meaning that
we have explicit upper bounds on all winding numbers for the first two
branes in the configuration).  

For the third brane, the  TCSB (\ref{eq:TCSB}) for the $P$ and $Q$
tadpoles shows that we have that either $P_3<T$ or $Q_3<T$.  Without loss
of generality we take $P_3<T$.  Thus, for the third brane $P_3$ and
$S_3$ are now constrained.  Using SUSY between branes $1$ and $3$ we
get that $\frac{R_1}{P_1} > \frac{R_3}{P_3}$.  Thus $R_3 < P_3
\frac{R_1}{P_1}$ and so $R_3$ is constrained.

We have thus determined upper bounds on all winding numbers for the
three branes (\ref{eq:3-branes}).  We now move on to the second step,
which involves finding the unique set of moduli consistent with
supersymmetry for the first three branes.  This will then allow us to
efficiently add to the first three branes all branes consistent with
the moduli.  Having three distinct branes in a configuration is a
necessary condition to uniquely determine the moduli, but not a
sufficient one (the system of three SUSY equations (\ref{eq:SUSY-1b})
for 3 moduli may not have a unique solution).  In order to actually be
able to solve for the moduli using the three branes
(\ref{eq:3-branes}), we need to prove that these three branes give
linearly independent constraints on the moduli.  Suppose that the
constraints are dependent, so that there exist an $\alpha, \beta$ such
that
\begin{equation}
\alpha(-\Ov{P_1},\Ov{Q_1},\Ov{R_1},\Ov{S_1}) +\beta (\Ov{P_2},-\Ov{Q_2},\Ov{R_2},\Ov{S_2}) = 
(\Ov{P_3},\Ov{Q_3},-\Ov{R_3},\Ov{S_3})
\end{equation}

From the linear relation on the first element of the vector we see
that either $\alpha<0$ or $\beta>0$, from the relation on the second
element we need $\alpha >0$ or $\beta <0$, from the third either
$\alpha <0 $ or $\beta <0$, and from the fourth, $\alpha>0$ or
$\beta>0$.  There are no solutions to this set of sign constraints.  

Having uniquely determined the moduli we can efficiently find all
branes consistent with this moduli so we can add them to the branes
(\ref{eq:3-branes}) in all ways compatible with the total tadpole
constraints.  In \ref{sec:space-models} we summarized a simple
argument showing that there are a finite number of such possible
configurations, and that the winding numbers for each brane can be
bounded.
For example, each brane with negative $P$ tadpole has positive tadpole
contributions $Q, R, S$ bounded by
\begin{equation}
\frac{Q}{j}, \; \frac{R}{k}, \;\frac{S}{l}< \gamma =
-\frac{P}{h}+\frac{Q}{j}+\frac{R}{k}+\frac{S}{l} \leq T(\frac{1}{h} + \frac{1}{j} + \frac{1}{k} + \frac{1}{l})
\end{equation}
Similar bounds can be given for branes with positive $P$ tadpole.
This bounds all winding numbers on additional branes to be added to
(\ref{eq:3-branes}) once the moduli are fixed.  Since all branes
contribute a positive amount
$\gamma_a> 0$ (by (\ref{eq:SUSY-2})) to the sum 
\begin{equation}
\sum_{a}\gamma_a \leq
 T(\frac{1}{h} + \frac{1}{j} + \frac{1}{k} + \frac{1}{l})
\end{equation}
we can combine the additional branes at fixed moduli in only a finite
number of ways compatible with the tadpole constraints, which are
easily enumerated.  This gives us a systematic way of constructing all
possible \a-brane configurations of type $[p,q,r,0]$.

\subsection{Results}
\label{sec:a-results}

We have performed a full search for all possible \Ab-brane
configurations using the algorithm described in the previous
subsection.  We find a total of 99,479 distinct configurations (with
no tilted tori), after removing redundancies from the permutation
symmetries on tadpoles and branes.  In Table 1 we show the
distribution of the number of \a-branes in these configurations.  Note
that with more than 3 \a-branes, the number of possible configurations
decreases sharply.  The configurations computed here are those which
satisfy the SUSY constraints for a common set of moduli and which
undersaturate the tadpole constraints.  
As we discuss in the following sections,
to form complete models associated with valid string vacua, \b-branes
compatible with the SUSY equations
and ``filler'' \c-branes must generally be added 
to any particular \a-brane
configuration to saturate the tadpole constraints, and then K-theory
constraints must be checked.  

\begin{center}
\begin{table}[ht]
\begin{center}
\begin{tabular}{|c |cccccccc |}
\hline
$n$ & 1 & 2 & 3 & 4 & 5 & 6 & 7 & 8\\
\hline
Number Configurations &$226$ &$30,255$ &$57,651$ &$9,315$ &$1,615$ &$361$ &$55$ &$1$\\
\hline
\end{tabular}
\caption{Number of configurations with $n$ \Ab-branes.}
\label{t:na}
\end{center}
\end{table}

\begin{table}[ht]
\begin{center}
\begin{tabular}{|l || *{4}{c|} }
\hline
\verb+p\q+ & 0    & 1             & 2    & 3 \\ \hline\hline
1     & 226  & 28560  & -    & - \\ \hline
2   & 1695 & 52761  & 3286 & - \\ \hline
3   & 857  & 5048   & 694  & 51 \\ \hline
4   & 105  & 689    & 170  & 0 \\ \hline
5   & 9    & 89     & 27   & 0 \\ \hline
6   & 2    & 12     & 0    & 0 \\ \hline
7   & 1    & 0      & 0    & 0 \\ \hline
8   & 1    & 0      & 0    & 0\\
\hline
\end{tabular}
\caption{Number of configurations with $p$ \a-branes with negative $P$
tadpole, $q$ \a-branes with negative $Q$ tadpole, and no branes with
negative $R$ or $S$ tadpole.}
\end{center}
\label{t:pq}
\end{table}

\end{center}

The number of \a-branes with negative $P$ and $Q$ tadpoles in
configurations of types $[p, 0, 0, 0]$ and $[p, q, 0, 0]$ is tabulated
in Table 2.  So, for example, of the 57,651 combinations with
3 \a-branes,  857 are of type [3, 0, 0, 0], 52,761 are
of type [2, 1, 0, 0], and the remaining 4033 are of type [1, 1, 1, 0].

While there are only 226 individual \a-branes which alone satisfy all
tadpole constraints, many more distinct individual \a-branes are
possible in combination with other \a-branes.  The number of distinct
(up to symmetry) \a-branes appearing in any configuration is 3259.  A
simple consequence of the Two Column SUSY Bound (\ref{eq:TCSB}) is
that no individual \a-brane can have more than one tadpole $> T$.
It is possible, however, to have an \a-brane with one large positive
tadpole, compensated by a negative tadpole on another brane.  The most
extreme case of this is realized in a two \a-brane combination in
which one brane has a tadpole $P = 800$
\begin{equation}
(-792, 3, 3, 88)_{(3, 1; 3, 1; -88, -1)} \; +
\; (800, 2, 5, -80)_{(2, 1; 5, 1; 80, -1)} \,.
\end{equation}

One of the most significant features of the \a-brane combinations
tabulated in Table~\ref{t:na} is that
\vspace*{0.05in}

{\it Any combination of \a-branes has at most one negative total tadpole}
\vspace*{0.05in}

This was
proven in \cite{Douglas-Taylor} for a combination of two \a-branes but
it is straightforward to generalize to any number of \a-branes.
Consider for example the tadpoles $P, Q$.
As in the discussion in \ref{sec:space-models}, for every brane
with negative $P_a = -P$ we have from the SUSY condition $Q_a/j  -
P/h > 0$, and for each brane with negative $Q_b = -Q$ we have
$-Q/j + P_b/h > 0$.  Thus, in the sum over all branes in any
configuration we have
\begin{equation}
\sum_{a} \frac{P_a}{h}  + \frac{Q_a}{j}  > 0
\end{equation}
so that only one of the total tadpoles $P, Q$ can be negative.  The same
holds for any pair so, as stated above, at most one total tadpole can be negative for
any combination.

\begin{figure}
\includegraphics[width=2.6in]{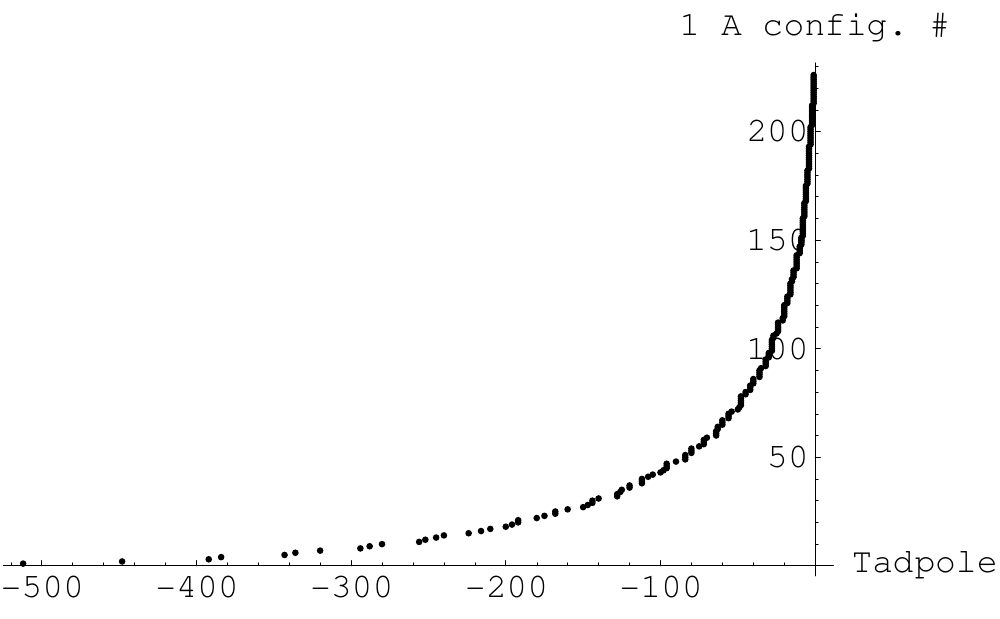}
\includegraphics[width=2.6in]{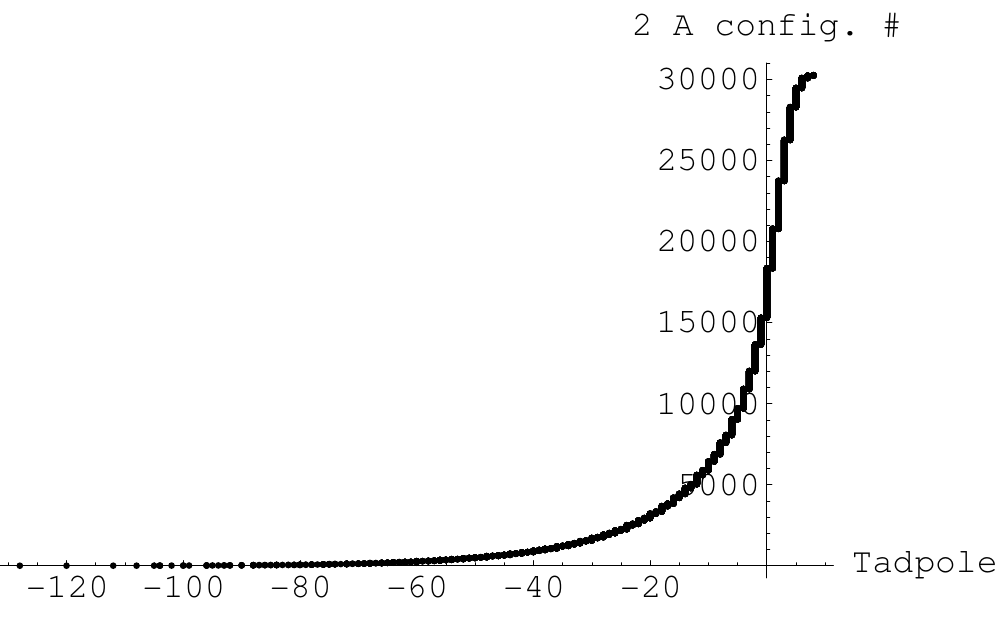}
\vspace*{0.1in}

\includegraphics[width=2.6in]{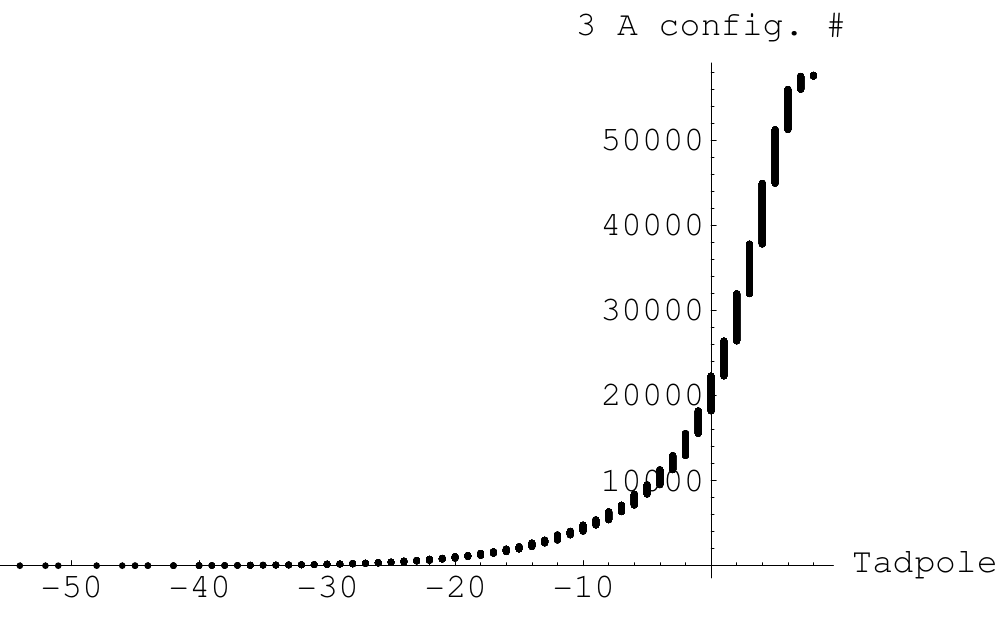}
\includegraphics[width=2.6in]{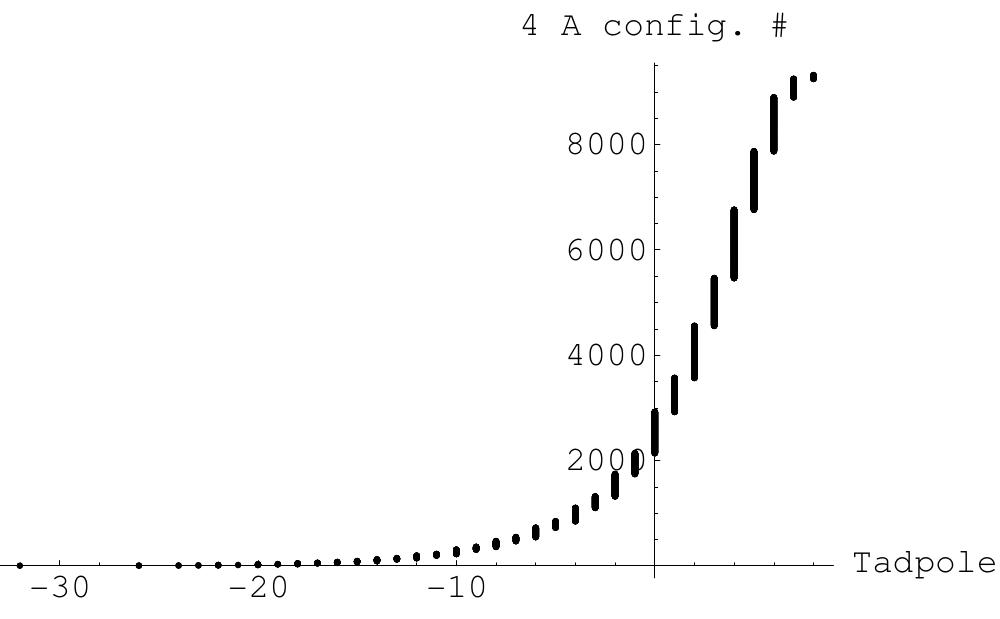}
\caption[x]{\footnotesize Minimum total tadpole in each configuration
  of 1, 2, 3, and 4 \a-branes (untilted tori).}
\label{f:a-tadpoles}
\end{figure}

As a consequence of this result, any combination of
\a-branes acts in a similar fashion to a single \a-brane.
Furthermore, when \a-branes are added, since the individual winding
numbers on each brane must be smaller, the maximum achievable negative
tadpole decreases quickly as the branes are combined.  Thus, a single \a-brane
can achieve the most negative tadpole.  Indeed, the single \a-brane
with the most negative tadpole (and no tadpoles $> T$) is
\begin{equation}
(-512, 8, 8, 8)_{(8, 1; 8, 1; -8, 1)} \,.
\label{eq:maximum-a}
\end{equation}
The combination of two \a-branes with the most negative total tadpole is
\begin{equation}
2 \times
(-64, 4, 4, 4)_{(4, 1; 4, 1; -4, 1)} 
=(-128, 8, 8, 8) \,.
\end{equation}
For three \a-branes, the most negative total tadpole is -54, and for
four \a-branes the most negative total tadpole is -32.  The
distribution of the smallest total tadpole for all combinations of 1-4
\a-branes (with no total tadpoles $> T$) is depicted in
Figure~\ref{f:a-tadpoles}.  In each case, the branes are ordered by
minimum total tadpole and distributed linearly along the vertical axis.

We thus see that most multiple \a-brane configurations have similar
properties to a single \a-brane with a minimum tadpole which is fairly
small in absolute value.  Only for combinations with a small number of
\a-branes are there configurations with substantially large negative
tadpoles.  Note, however, that although multiple \a-brane
configurations act similar to a single \a-brane in terms of the total effect
on tadpole contributions, they may act very differently when it comes
to satisfying K-theory constraints.

In constructing more general models with multiple stacks of D-branes,
the greatest variety of constructions is possible when the smallest
total tadpole coming from the \a-brane sector is as negative as
possible.  In particular, the greatest flexibility in adding \b-branes
and \c-branes is afforded when the \a-brane sector has a very negative
total tadpole and the positive total tadpoles are also small.  This is
realized primarily for single \a-branes with very negative tadpoles.
There are 33 single \a-branes with negative tadpole $\leq -128$, and
71 single \a-branes with negative tadpole $\leq -54$.  A typical
example is the \a-brane
\begin{equation}
(-168, 3, 7, 8)_{(3, 1; 7, 1; -8; -1)} \,.
\label{eq:a-example}
\end{equation}
There are also 403 combinations of two \a-branes with negative tadpole
$\leq -54$.  The two \a-brane combinations have larger positive total
tadpoles than the single brane configurations with similar negative
tadpole, so that the single brane gives space for a wider variety of
added \b- and \c-branes.  As we shall see, most of the diversity of
models with multiple brane stacks comes from these single (and some
double) brane configurations with highly negative tadpoles.  These
configurations are also associated with relatively large integer
moduli $h, j, k, l$, as we discuss in more detail in Section
\ref{sec:comparison}.  On the other hand, as we discuss in Section
\ref{sec:K-theory}, the K-theory constraints become more restrictive
for single \a-branes with very large negative tadpoles.  This
effect mitigates to some extent the role of the single \a-branes with
extremely negative tadpoles in generating diversity in constructions
of interesting physics models.

Finally, we discuss the question of \a-brane combinations with tilted
tori.  As described in Section \ref{sec:t6}, on a tilted torus the
tadpole and winding number conditions are very similar to those on a
rectangular torus, but the winding numbers on the tilted torus must
satisfy $n_i \equiv 2 \tilde{m}_i$ (mod 2).  The relative primality
condition is weakened so that $n_i$ and $2 \tilde{m}_i$ can both be
even, if $n_i/2 \not\equiv \tilde{m}_i$ (mod 2).  We can realize all
brane combinations realizing these constraints by simply constructing
all brane combinations for the rectangular torus, relating $m_i$ in
the construction to $2 \tilde{m}_i$ and imposing the additional
condition that on a tilted torus there must be an even number of any
brane with $n_i \not\equiv m_i$ (mod 2), corresponding to half that
number of branes with twice the $n_i$ and $\tilde{m}_i$ equal to that
$m_i$.  (For several tilted tori, each tilted direction in which $n_i
\not\equiv 2 \tilde{m}_i$ requires that we double the effective number
of branes in the counting for rectangular tori to get a single brane
on the tilted tori).  Note that when a subset of tori are tilted, the
permutation symmetry on tadpoles is broken.  In particular, this means
that a brane configuration on rectangular tori may give several
configurations with 1 or 2 tilted tori, depending on which torus/tori
is/are tilted.  For example, the single \a-brane of
(\ref{eq:a-example}) is a valid \a-brane (using $m_i \rightarrow 2
\tilde{m}_i$ on the tilted tori)
if either the first or second
torus is tilted, but not if the third torus is tilted since $n_3\not
\equiv 2 \tilde{m}_3$ (mod 2).  If we are counting all \a-brane
combinations with the first torus tilted, then, after using the
permutation symmetry (\ref{eq:a-example}) gives the allowable
\a-branes
\begin{equation}
(-168, 3, 7, 8)_{(3, \tilde{1/2}; 7, 1; -8; -1)}, \;\;\;\;\;
(-168, 7, 3, 8)_{(7,  \tilde{1/2}; 3, 1; -8; -1)} 
\label{eq:a-example-tilt}
\end{equation}
where the tilde denotes winding numbers $\tilde{m}_i$ on the tilted torus.
We have computed the number of \a-brane configurations with 1, 2, and
3 tilted tori in this fashion.  The results are given in
Table~\ref{t:tilted-a}.

\begin{table}
\begin{center}
\begin{tabular}{|c || *{8}{c} |}
\hline
\verb+# tilted tori\n+ & 1 & 2 & 3 & 4 & 5 & 6 & 7 & 8 \\ \hline\hline
1  & 242 & 24783 & 27712 & 10068 & 1375 & 477 & 36 & 1 \\
2  & 136 & 5897     & 4868     & 3127     & 422    & 222 & 9     & 0 \\
3  & 29     & 471     & 277   & 354     & 38  & 36  & 0     & 0\\
\hline
\end{tabular}
\end{center}
\caption{Number of configurations of $n$ \a-branes having all total
  tadpoles $\leq T = 8$ with 1, 2, or 3
  tilted tori.}
\label{t:tilted-a}
\end{table}

\section{Models containing gauge group $G$}
\label{sec:fixed-gauge}

Using the set of all possible configurations of \Ab-branes, as
described in the previous section, it is possible to efficiently
generate all brane configurations which realize many features of
interest.  In particular, given any fixed gauge group $G$ it is
possible to construct all distinct brane combinations which realize
this gauge group as a subgroup of the full gauge group in any model in
polynomial time.  In principle, construction of all SUSY models is
possible, but this is computationally intensive as the total number of
models is quite large.

There are several reasons for focusing on the problem of constructing
all realizations of a fixed group $G$ rather than simply enumerating
all models.  This approach significantly simplifies the computational
complexity, while still extracting some of the most interesting data.
From a purely model-building perspective, say one is interested in
constructing all standard-model like brane configurations.  For a
given realization of the group $G_{321} = SU(3) \times SU(2) \times
U(1)$ in terms of a set of 3 + 2 + 1 branes, there may be a large
number of ways of completing the configuration to saturate the tadpole
equations.  But much of the physics of the model, such as the number
of generations of ``quarks'' carrying charge under $SU(3)$ and $SU(2)$
depends only on the choice of branes to realize $G$ and is independent
of the way in which this model is completed with extra branes.  The
extra branes may generate a hidden sector or chiral exotics which are
of interest, but it is probably more efficient for model building
purposes to first consider all realizations of $G$, and then to
explore the possible extra sectors only of those realizations which have
physical properties of interest.

From a more general point of view, the purpose of constructing all
configurations with fixed gauge subgroup $G$ is to get a clear handle
on what the important factors are which control the distribution of
models.  By considering the variety of ways in which a gauge subgroup
like $G = SU(3) \times SU(2)$ can be realized in any SUSY IBM model
on $\orbifold$, for example, we gain insight into the mechanism
responsible for generating the bulk of these configurations.  
This also provides a clear way to
analyze more detailed features of these constructions such as the
number of generations of matter fields in various representations.

In the first part of this section (Subsection \ref{sec:general-subgroup}) we
describe the general method of computing all brane configurations
which generate a gauge subgroup $G$; we then explicitly compute all such
configurations for gauge groups $U(N)$ in Subsection \ref{sec:un}
and
$SU(3) \times SU(2)$ 
in
Subsection \ref{sec:32}.  By looking at the distribution of these
gauge groups and associated tadpoles, we gain insight into how the
diversity of realizations of these groups is associated with \a-branes
with large negative tadpoles, as well as providing useful tools for
model building.
We also construct all brane configurations realizing $SU(N) \times
SU(2) \times SU(2)$ in \ref{sec:n22}.
In \ref{sec:K-theory} we discuss the K-theory constraints and how they
can reduce the total number of allowed realizations of any fixed $G$.
Finally, in subsections
\ref{sec:models} and
\ref{sec:other-orbifolds} we relate the results described here to
earlier work on IBM model building on this and other toroidal orbifold models.

\subsection{Systematic construction of brane realizations of $G$}
\label{sec:general-subgroup}

As mentioned above, the problem of finding all ways in which a fixed
group $G$ can be realized as a subgroup of the full gauge group can be
solved in a straightforward way in polynomial time given the results
of the Section \ref{sec:a}.  Any complete model containing a set of
branes individually satisfying the SUSY constraints for a common set of moduli and collectively
solving the tadpole and K-theory constraints, contains either no \a-branes or some given set
of \a-branes which must be one of the 99,479 configurations enumerated
above.  Given a configuration of \a-branes, there is a finite number
of ways in which \b- and \c-branes can be added to saturate the
tadpole conditions, since the \b- and \c-branes have only positive
tadpoles.  Thus, to determine all realizations of $G$, we just need to
run through each of the roughly $10^5$ possible \a-brane combinations and
for each determine all realizations of $G$ through adding \b- and
\c-branes to the given \a-brane combination without oversaturating the
tadpole constraints.

To be explicit, say we want to find all models containing the gauge
group $G =SU(N_1) \times SU(N_2) \times ...  \times SU(N_r)$.  Each of
the $r$ stacks can be made up of \Ab\ , \Bb\, or \Cb \ type branes (for
\a- and \b-branes $SU(N_i)$ would be realized as a subgroup of
$U(N_i)$, while for \c-branes, $SU(N_i)$ would be realized as a
subgroup of a symplectic group $Sp(N)$ as discussed in more detail
below).

The first step in explicitly constructing all realizations of $G$ is
looping through our list of all \Ab-brane configurations.  For each
\Ab-brane configuration we see if there are $N_i$ duplicates of any
brane.  If so, then these $N_i$ \Ab-branes can provide a factor of $U(N_i)$
to the gauge group.  We form all possible combinations of branes in
the \Ab-brane configuration which can be used to compose parts of the
group $G$, with the remaining parts arising from extra branes which
must be added to the model.  For each of
these realizations of a subgroup of $G$ by some branes in a
configuration of \a-branes, we then consider all possibilites of \Bb\
and \Cb\ type branes that can be added in stacks to fill out the
remaining needed components of $G$. 

Algorithmically, adding \Bb-branes to the configuration of \a-branes
is straightforward.  Since we have already included all the \Ab-branes
that will go into the configuration, and since \Bb-branes have only
positive tadpoles, all the tadpoles will be bounded by the tadpole
constraint.  Thus, to find all ways of including a stack of \b-branes
which are compatible with a given \a-brane combination, we proceed by
scanning over possible winding numbers of the \Bb-brane to be added
that are consistent with the tadpole constraints. Once a compatible stack of $N$
\Bb-branes is found, we check to see if this \Bb-brane along with the
branes already in the configuration satisfy the SUSY condition, by
confirming that the resulting constraints on moduli are compatible.  All possible ways of adding \b-brane stacks within the
tadpole constraints can be constructed in this fashion.  The
addition of stacks of \c-branes is even simpler, since there are only
four different \Cb-branes that can be added and the \c-branes do not
affect the SUSY conditions.

After constructing the set of all realizations in this way, we may
have multiple instances of the same realization, for example
associated with different extra \a-brane configurations.  To reduce
the final set of configurations to a single instance of each
equivalent realization, we must drop the extra branes, put each
configuration in some canonical form, and drop copies. 
Note that in this class of brane configurations we have a clear criterion for
determining equivalence of solutions, using the symmetries described
in Section \ref{sec:t6}, unlike for example the situation described in
\cite{Dienes-equivalent}. 

Thus, in a straightforward way we can scan over all possible
inequivalent ways of building the gauge group $G$ from \a-, \b-, and
\c-brane stacks in a way which is compatible with the supersymmetry
and tadpole constraints.  Note that we are not checking the K-theory
constraints at this stage, since we are only generating a subset of
the complete set of branes in any given model.  Thus, the set of
realizations generated through this algorithm may be over-complete.
While generically additional branes can be added in many ways, some of
which will satisfy the K-theory constraints, in some cases,
particularly when our realization of $G$ comes close to satisfying the
tadpole constraints, there may be no complete model containing this
realization which satisfies the K-theory constraints.  This must be
checked in a case-by-case fashion for models of interest.

We have so far concentrated on the case when the tori are untilted.
For tilted tori we proceed as discussed above for enumerating \a-brane
stacks.  We only keep configurations where there are an even number of
branes with $n_i\not\equiv m_i = 2 \tilde{m}_i$ on the tilted tori,
noting that the resulting gauge group for a stack of $2^kN$ such
branes with $n_i \not\equiv m_i$ on $k$ tilted tori is $U(N)$.
Practically, we can find configurations realizing a desired gauge
group on a compactification with tilted tori by computing the
configurations on untilted tori, and then checking whenever there is
a
stack of $N$ branes with $n_i \not\equiv m_i$ on $k$ tilted tori
that an
additional $N(2^k -1)$ branes of this kind can be added without oversaturating
the tadpole conditions.  (For the \a-branes, we just need to confirm
that there are an additional $N (2^k -1)$ of these branes available in the
\a-brane combination used at the first step of the analysis.)
Clearly, this means that with more tilted tori there will be fewer
realizations of $G$.

\subsection{Realizations of $SU(N)$}
\label{sec:un}

As a simple example, we consider the construction of all possible
brane realizations of the group $SU(N)$ as a subgroup of the full gauge
group.  There are 3 ways in which the group $SU(N)$ can be realized.
\vspace*{0.05in}

\noindent
{\it i})
The group $SU(N)$ can be realized as a subgroup of the $U(N)$
associated with $N$ identical \a-branes\footnote{Note that when
  $SU(N)$ is realized as a subgroup of $U(N)$, as in {\it i}) and {\it
    ii}), the extra $U(1)$ factor
often becomes anomalous and gets a mass through the Green-Schwarz
mechanism \cite{afiru, imr}.}.
To identify all ways in which this can be done we just need to scan
over all $10^5$ known \a-brane combinations for configurations
including $N$ copies of the same brane, and then list all \a-branes
for which $N$ copies appear in some \a-brane combination.
\vspace*{0.05in}

\noindent
{\it ii}) The group $SU(N)$ can be realized, again as a subgroup of
$U(N)$, through $N$ identical \b-branes.  For each combination
$\alpha$ of \a-branes, we look at all possible ways in which a
\b-brane $\beta$ can be chosen so that combining $N$ copies of $\beta$
with $\alpha$ gives total tadpoles which are all $\leq T$.  This
condition puts strong constraints on the winding numbers of $\beta$.
For each $\beta$ which combines with $\alpha$ without exceeding the
tadpole constraints, a further check must be done that the system of
linear equations given by the SUSY equalities (\ref{eq:SUSY-1}) for
the branes in $\alpha$ and $\beta$ admit at least one solution.
Finally, symmetries must be considered so that only one example of
each such $\beta$ configuration is included in a final list.  (While
the same brane stack $\beta$ may be associated with many different
\a-brane combinations $\alpha$, these represent ``extra'' branes in
the same way as additional \b-branes or \c-branes completing the
tadpole constraints, so do not really realize distinct realizations of
$SU(N)$.)
\vspace*{0.05in}

\noindent
{\it iii}) The group $SU(N)$ can also be realized as a subgroup of
$Sp(N)$ arising from $N$ identical \c-branes.  Checking for this
possibility for each combination $\alpha$ of \a-branes is
straightforward; any total tadpole of $\alpha$ which is less than
$T-N$ can be associated with $N$ additional \c-branes.  The value $N =
2$ is a special case, since $SU(2) = Sp(1)$, so $SU(2)$ can be
realized from a single \c-brane.  Since there is really, up to
symmetry, only one possible stack of $N$ \c-branes, there is one
possible realization of each $SU(N)$ as a subgroup of $Sp(N)$ up to
the maximum of $N = 520$, which can be realized in the presence of the
\a-brane (\ref{eq:maximum-a}).  There is one more subtlety here
relevant for model-building.  Although the group $SU(N)$ can be
realized as a subgroup of $Sp(N)$ with $N$ \c-branes, the fields
transforming only under that $Sp(N)$ live in the antisymmetric
representation and cannot break $Sp(N)$ down to only $SU(N)$.  By
turning on bifundamentals, for example between two stacks of $N$
\c-branes with different tadpoles, the symmetry can be broken down to
$SU(N)$; this corresponds to brane recombination, giving a different
set of branes.  Alternatively, if we have $2 N$ \c-branes, these
branes can be moved away from the orientifold plane, giving a gauge
group $U(N) \supset SU(N)$, as described in \cite{cl3}.  Using
this mechanism, the largest $SU(N)$ which can be realized without the
remainder of an $Sp(N)$ is $SU(260)$ by 520 \c-branes in combination
with (\ref{eq:maximum-a}).

We have carried out the necessary computation for each type of brane.
The number of distinct realizations of $SU(N)$ in terms of \a-branes,
\b-branes, and \c-branes is shown in Table~\ref{t:un}.

\begin{table}
\begin{center}
\begin{tabular}{|c| r | r | r | r | r |r | r | r | r  | r |}
\hline
\verb+ type\N + & $U(1)$ &2 & 3 & 4 & 5 & 6 & 7 & 8 & $> 8$& $> 520$\\
\hline
\a  &3259 &250 &59 & 17&8 &3 & 1& 1&0 & 0\\
\hline
\b  &7067 & 1144 &377 & 151 &82 &39 &15 & 1 & 0&0\\
\hline
\c  &1 &1 & 1 &1 &1 &1 &1 &1 &1 & 0\\
\hline
\end{tabular}
\caption[x]{\footnotesize  Numbers of distinct ways in which $SU(N)$
  ($U(1)$ for $N = 1$)
  can be realized by \a, \b, or \c-type branes as a subgroup of the
  full gauge group (with untilted tori).}
\label{t:un}
\end{center}

\end{table}

Most of the realizations of $SU(N)$ arise from stacks of $N$
\b-branes.  To understand the origin of the numbers in this table more
clearly let us consider the case of $SU(7)\subset U(7)$ arising from 7
identical \b-branes.  Each \b-brane has two nonvanishing tadpoles, so
up to symmetries the \b-brane involved has tadpoles $(P, Q, 0, 0)$
with $P \geq Q > 0$, so the full $U(N)$ stack has total tadpole $(7P,
7Q, 0, 0)$.  Since every \a-brane combination has at most one negative
tadpole, we must have $Q = 1$.  This constrains the winding numbers
$n_1, m_2, m_3$ to be unity (with canonically chosen signs), so that
the only freedom is in winding numbers $n_2, n_3$ with $P = n_2 n_3$.
In the absence of any \a-branes, we can only have $P = 1$.  Any
\a-brane combination included must have total $Q = 1$, and $P < 0$.
Among single \a-branes, the one with the most negative tadpole and $Q
= 1$ is
\begin{equation}
\alpha_{64} = (-64, 1, 8, 8)_{(1, 1; 8, 1; -8, -1)} \,.
\label{eq:a64}
\end{equation} 
With more \a-branes, constrained to have total $Q = 1$, the negative
tadpole decreases rapidly in absolute value.  For two \a-branes, for
example, the most negative $P$ tadpole comes in the combination $(-9,
1, 7, 8) = (-24, 2, 3, 4)_{(2, 1; 3, 1; -4, -1)} + (15, -1, 5, 3)_{(1,
-1; 5, 1; 3, 1)}$.  A single \a-brane and a single \b-brane of this type are always compatible under SUSY. Thus, the set of possible \b-branes giving $U(7)$ is just the set of
$n_2, n_3$ with $7 n_2 n_3 \leq 72$, where we can choose $n_2 \geq n_3$
using the permutation symmetry between the second and third tori.
There is no relative primality constraint as these winding numbers are
on different tori.  There are precisely 15 such combinations, giving
the entry in the table above.

A similar story holds for $U(N)$ with smaller $N$ arising from a stack
of $N$ identical \b-branes.  For $N = 6$, the \a-brane can have $Q =
2$, which allows $P = -128$.  The number of $n_2 \geq n_3$ with $6n_2 n_3 \leq
136$ is 39, as in Table~\ref{t:un}.  The story is slightly more
complicated for $N \leq 4$, since then the \b-brane can have $Q > 1$,
but a similar analysis shows that precisely the number of \b-branes
indicated in Table~\ref{t:un} can be included $N$ times in a model,
almost always with a single additional \a-brane.

\begin{figure}
\begin{center}
\includegraphics[width=4in]{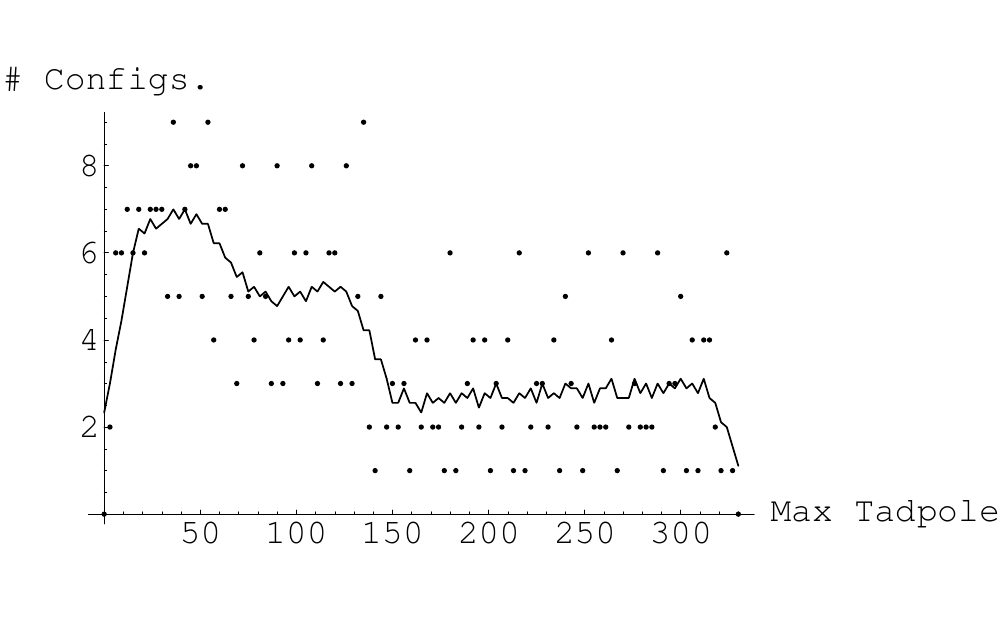}
\end{center}
\caption[x]{\footnotesize 
Distribution of the
maximum tadpole for the 3 branes forming
  each of the 437 distinct $SU(3)$ realizations.  Solid curve is data
  averaged over window including the nearest 4 data points in each direction for
  clarity (untilted tori).}
\label{f:n3}
\end{figure}

This analysis gives the first clear example of one of the primary
conclusions of this paper: the greatest diversity of
configurations realizing a particular feature (such as a fixed
subgroup of the full gauge group) arises in association with one or
sometimes more ``extra'' \a-branes. These \a-branes lie in the tail of the
distribution and give a very negative tadpole
allowing ``phase space'' for the range of winding numbers.  Indeed,
most of the \b-brane stacks giving rise to the range of $SU(N)$
configurations have a single large positive tadpole, requiring a
single extra \a-brane with a very negative tadpole.
A simple way to see this is to graph the distribution of
the largest positive
tadpole associated with the stack of $N$ branes giving the $SU(N)$
gauge group.  The distribution of maximum total tadpoles for the
3 branes giving the $SU(3)$
in the case $N = 3$ is graphed in Figure~\ref{f:n3}.
For these configurations the maximum tadpole ranges 
from 3 through 327 (this number includes the factor of 3 from the
stack of 3 branes).  Most of these tadpoles are quite large compared
to the negative tadpole of a typical \a-brane combination.  Indeed,
over 75\% of the $SU(3)$ realizations have a maximum tadpole
contribution above 50, and can only be realized with single \a-branes
in the tail of the distribution shown in Figure~\ref{f:a-tadpoles}.
The distribution in Figure~\ref{f:n3}  is rather discontinuous due to
the small numbers of configurations involved.  

A statistical analysis of models containing the group $SU(5)$ was
carried out in \cite{Gmeiner-Stein}.  They found almost 7000 models
containing the group $SU(5)$ at small moduli.  This number is much
greater than the number 91 in Table 4 giving the number of distinct
realizations of $SU(5)$ since they include all possible configurations
of additional branes saturating the tadpole condition to form complete
models.  In fact, most of the 7000 models they produced most likely
come from a small subset of the 91 distinct $SU(5)$ configurations,
since those with large winding numbers produce larger moduli.  We
discuss related issues in more detail in Section \ref{sec:comparison}.

\subsection{Realizations of $SU(3) \times SU(2)$}
\label{sec:32}

As a somewhat more complex example we have computed all realizations
of  $G =SU(3) \times SU(2)$.  As described in Subsection \ref{sec:general-subgroup}
we can systematically find all such realizations by first considering
all \a-brane combinations, finding all ways one or both components of
the gauge group can be included in the \a-brane combination, and then
adding stacks of \b- or \c-branes to complete the group $G$, including
cases where both the $SU(3)$ and the $SU(2)$ come from \b-\ or \c-branes.
This group is of obvious phenomenological interest, as every
standard-model like construction in this framework must contain at a
minimum the group $G = SU(3) \times SU(2)$ as a subgroup.  So finding
all possible constructions of this group represents a step towards
identifying all realizations of standard-model like physics in any
class of compactifications.

\subsubsection{Enumeration of distinct realizations for $SU(3) \times SU(2)$}

There are 9 different ways in which the different types of branes can
be combined to form the group $G$.  The numbers of ways in which this
can be done (without tilted tori) are tabulated in Table~\ref{t:32}.
Note that we are only requiring a single \c-brane for $SU(2)= Sp(1)$,
and including configurations with 3 \c-branes where $SU(3) \subset
Sp(3)$, although as discussed in \ref{sec:un} additional branes are
needed to allow the breaking of $Sp(N)$ to just $SU(3)$.
(Configurations where 6 \c-branes realize $SU(3)$, so that the
reduction $Sp(6)\rightarrow SU(3)$ is possible by brane splitting, are
indicated in parentheses.)  Thus, we are solving the mathematical
problem of finding all realizations of $SU(3) \times SU(2)$ in any way
that it can be realized as a subgroup of the full gauge group
(including when the branes are the same, giving a group $SU(5)$).
This includes any possible configuration which would lead to a model
containing any extension of the standard model, but not every
configuration constructed here will correspond to a full model with
explicit $SU(3) \times SU(2)$ gauge subgroup.  Each of the
configurations found here can be extended in one or more ways to a
full model, for which the K-theory constraints must be checked.

\begin{table}
\begin{center}
\begin{tabular}{ | c | r | r | r |}
\hline
3 \verb+\+ 2 &\a &\b &\c\\
\hline
\a & 84 & 939 & 169\\
\b & 1802 & 164057 & 1274\\
\c & 554 (316) & 2774 (1595) & 2 (2)\\
\hline
\end{tabular}
\end{center}
\caption[x]{\footnotesize Numbers of distinct ways in which $G = SU(3)
 \times SU(2)$ can be realized through combinations of \a-, \b-, and
 \c-branes (untilted tori).  Numbers in parentheses are for $SU(3)
 \subset Sp(6)$.}
\label{t:32}
\end{table}

Of the total of 171,655 constructions of $G$, by far the greatest
number arise from combinations of \b-brane stacks.  Again, in most
cases there is a single extra \a-brane needed to push down the total
tadpole.  In the presence of this \a-brane there is a tradeoff between
\b-branes and \c-branes in the ``phase space'' of brane
configurations.  When $T$ is large there are many more \b-branes
than \c-branes possible within the tadpole limits, while when $T$ is small,
\c-branes contribute smaller tadpoles and are easier to add.  In this
situation, while $T = 8$ is reasonably small, the wider range of
possibilities for \b-branes wins out and these provide the widest
range of possible realizations of the desired gauge group.
Of the more than 164,000 combinations of 3 \b-branes with one set of
winding numbers and 2 \b-branes with another set of winding numbers,
all but 2 configurations can be realized with either no \a-branes or a
single \a-brane.  These two exceptional cases are given by
\begin{eqnarray*}
3 \times (0, 4, 1, 0)_{(2, 1; 1, 2; 0, -1)}&+ &
2 \times (0, 0, 1, 2)_{(0, -1; 1, 2; 1, 1)}\\
3 \times (2, 0, 1, 0)_{(2, 1; 1, 0; 1, -1)} & + & 
2 \times (0, 0, 1, 8)_{(0, -1; 1, 8, 1, 1)}
\end{eqnarray*}
Each of these two cases requires two additional identical \a-branes,
$2 \times (4, -2, 1, 2)_{(2, -1; 1, 1; 2, 1)}$ in the first case, and
$2 \times (1, 4, 1, -4)_{(1, 1; 1, 4; 1, -1)}$ in the second.  The integer moduli
are uniquely fixed in both cases, to $(h,j,k,l)= (8, 1, 4, 4)$ and $(h,j,k,l)= (8, 1, 16,
1)$ respectively.

\begin{figure}
\begin{center}
\includegraphics[width=4in]{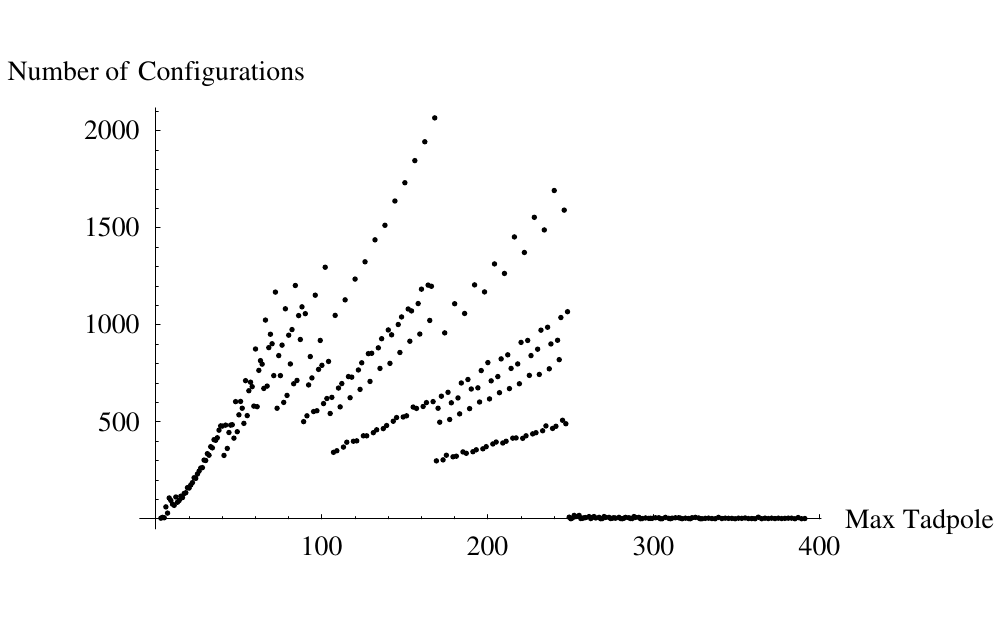}
\end{center}
\caption[x]{\footnotesize Number of distinct $SU(3) \times SU(2)$
  configurations of \b/\c-branes (on untilted tori)
with given maximum total tadpole.
  Form of distribution shows most configurations
  are only possible with addition of an extra \a-brane with a large negative
  tadpole.}
\label{f:32-tadpoles}
\end{figure}

As in the $SU(N)$ case discussed in the previous subsection, it is
helpful to graph the distribution of maximum total tadpole to get a
sense of the distribution of models.  In Figure~\ref{f:32-tadpoles} we
graph the number of realizations of $SU(3) \times SU(2)$ with
different maximum total tadpole contributions (including only
configurations composed of \b- and \c-branes).  As in the $SU(3)$
case, most of these tadpoles are only compatible with a single extra
\a-brane with a very negative tadpole in the tail of the distribution.
As discussed above, the sets of realizations computed here are
overcomplete, since many of these configurations cannot be completed
to complete models compatible with the K-theory constraints.  We
describe this reduction in more detail in Subsection \ref{sec:K-theory}.
\vspace*{0.1in}

While the focus of this paper is not on realistic model building, and
is rather on developing general methods
and a systematic understanding of the space of models on the
$\orbifold$ orientifold, it is interesting to push this construction
slightly further in the model-building direction to get further
information about the distribution of models in this framework.  To
this end, we have considered increasing the gauge group to $SU(3)
\times SU(2) \times U(1)$; we have also studied the distribution of
``quark'' generation numbers in the complete set of $SU(3) \times
SU(2)$ realizations just described.  We briefly describe these further
studies in the remainder of this subsection.

\subsubsection{Gauge group $SU(3) \times SU(2) \times U(1)$}

\begin{figure}
\begin{center}
\includegraphics[width=4in]{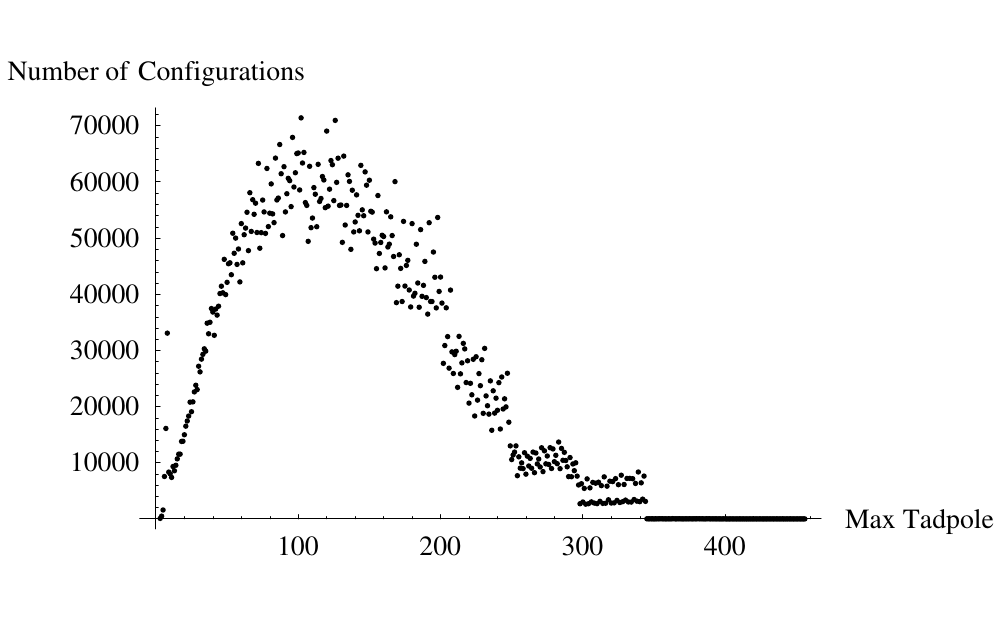}
\end{center}
\caption[x]{\footnotesize Number of distinct $SU(3) \times SU(2)
  \times U(1)$ brane configurations (on untilted tori) with given maximum total tadpole.
  Form of distribution with peak around 100 shows most configurations
  are only possible with addition of an extra \a-brane with a large negative
  tadpole.}
\label{f:321}
\end{figure}

We have generated all realizations of the larger gauge group $G_{321} =
SU(3) \times SU(2) \times U(1)$, where the $SU(3) \times SU(2)$
brane configurations described above are extended with one more brane
of either \a, \b, or \c-type ({\it i.e.}, the $U(1)$ does not come
solely from $U(3)$ and/or $U(2)$ groups in which the $SU(3)$, $SU(2)$
are embedded).  For the case when none of the tori are tilted, there
are approximately 13.7 million distinct ways of realizing this larger
gauge group.  About half of these configurations come from having a 3-stack of
\c-branes with a 2-stack of \b-branes and a 1-stack of \b-branes, along
with 1 extra \a-brane (we can write this as 3C $\times$ 2B $\times$ 1B
with extra A).  The majority of the other half of the configurations come from
an extra A-brane with 3B $\times$ 1C $\times$ 1B (with $SU(2) = Sp(2)$) or 3B $\times$ 2B $\times$ 1B, or 3B $\times$ 2B
$\times$ 1A.
As discussed above, the balance between \c-branes and \b-branes in the
construction depends on the tadpole $T$, which in this case is at an
intermediate point where the smaller contribution of \c-branes
competes well with the larger number of possible \b-branes in forming
the $SU(3)$ part of the group.
As in the case of $SU(3) \times SU(2)$, we graph the maximum total tadpole
from each brane configuration forming $SU(3) \times SU(2) \times
U(1)$.  The number of configurations with a given maximum total tadpole peaks
around a tadpole of 100, indicating that almost all of these configurations
depend upon an extra \a-brane with a highly negative tadpole.
Note that the large number of configurations involved gives the graph in
Figure~\ref{f:321} a much smoother appearance than the corresponding
graphs for $SU(3)$ or $SU(3) \times SU(2)$ depicted in
Figures~\ref{f:n3},~\ref{f:32-tadpoles}, although the discrete nature
of the constraint problem becomes apparent at larger values of the
tadpole.
As above, the configurations graphed in Figure 4 do not represent
complete models.

\subsubsection{Tilted tori}

We have carried out the computation of the number of distinct  brane
combinations satisfying SUSY and tadpole constraints and realizing groups
$SU(3) \times SU(2)$ and $ SU(3) \times SU(2) \times U(1)$ (again,
with an independent $U(1)$) for any number of tilted tori from 0
through 3.  The number of configurations for the four cases is given in Table 
\ref{t:321table}.
\begin{table}[ht] 
\begin{center}
\begin{tabular}{|c |cccc |}
\hline
Number Tilted Tori & 0 & 1 & 2 & 3\\
\hline
Number 3-2 Configurations &$171,655$ & $44,658$ & $2,026$ & $40$ \\
Number 3-2-1 Configurations  &$13,724,917$& $2,406,352$ & $103,652$ & $1,222$\\
\hline
\end{tabular}
\end{center}
\caption{Number of 3-2 and 3-2-1 configurations.}
\label{t:321table}
\end{table}

As the number of tilted tori is increased, the number of models
decreases substantially.  As discussed previously, this makes sense
since we can effectively treat models on tilted tori as a subset of
models on rectangular tori, where branes with opposite parity winding
numbers on tilted tori must appear even numbers of times.

The dramatic increase in the number of possible configurations when an extra
$U(1)$ brane is included illustrates the rapid combinatorial growth of
the total number of models when additional branes are included.
Indeed, typical 3-2-1 configurations are compatible with ${\cal O}
(10)$ distinct combinations of extra \a-branes.  In the enumeration of
such configurations, which began with distinct \a-brane configurations
and added branes in all ways to realize $G_{321}$, before we deleted
duplicate configurations to obtain the 13.7 million distinct 3-2-1
configurations, there were ${\cal O} (10^8)$ configurations
including \a-brane information.

For any specific realization of $G = SU(3) \times SU(2)$, all possible
combinations of extra branes which saturate the tadpole condition could
be determined.  In principle this could be done for all models, but it
would lead to hundreds of millions, probably billions of total models,
so a fairly extensive computer project would be involved.  Since much
of the relevant physics (such as the matter content in the
bifundamental representations of the group $G$)
can be determined just from the construction of $G$, it seems more
pragmatic to approach any systematic attempt to model building by
first constructing the gauge group of interest, and then isolating the
subset of models with further desired properties.  After this, the set
of extra branes giving rise to additional gauge groups and hidden or
exotic matter can be systematically determined.

\subsubsection{Distribution of generation numbers}

In \cite{Gmeiner, Douglas-Taylor}, a variety of models were considered
and correlations between gauge group and generation numbers were
studied.  In general, it was found that there was no strong
correlation between gauge group and generation numbers.  We have
explicitly computed the numbers of generations of ``quarks'' which
transform in the fundamental representation of $SU(3)$ and the
antifundamental (which is equivalent to the fundamental)
representation of $SU(2)$, for the various brane configurations
tabulated in Table~\ref{t:32}.  As discussed in Section
\ref{sec:t6}, only when there is at least one tilted torus can there
be an odd intersection number between branes which are not \c-branes.

\begin{figure} \label{IntNumFig}
\begin{center}
\includegraphics[width=4in]{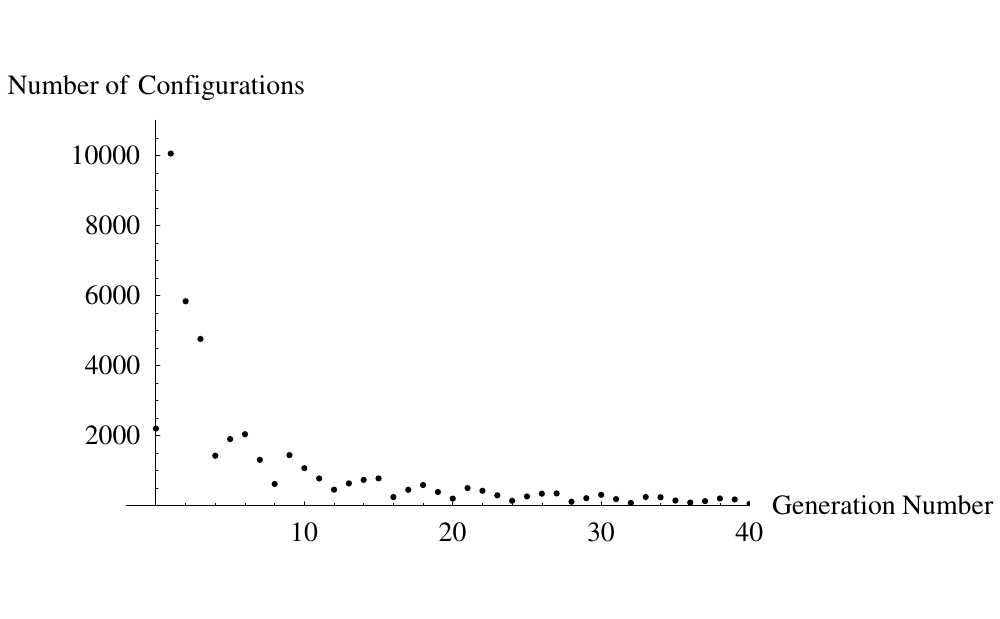}
\end{center}
\caption{Distribution of number of ``quark'' generations for  $SU(3)
  \times SU(2)$ configurations
with a single tilted torus.}
\label{f:intersection}
\end{figure}

We have analyzed the intersection numbers of all the configurations
described above giving $SU(3) \times SU(2)$ in the case of one tilted
torus.  The sum of intersection numbers $I_{32}+I_{32'}$, which gives
the number of ``quark'' generations, is computed for all 44,658 of
these configurations (using only $I_{32}$ when the $SU(2)$ comes from
a \c-brane), and the resulting distribution is graphed in
Figure~\ref{f:intersection}.

The number is peaked at 1, and drops off fairly rapidly, with
typical generation numbers of order ${\cal O} (10)$.  There is no
particular enhancement or suppression of 3 generations; 4760, or about
10\% of the configurations have 3 generations of ``quarks''.
Thus, 3 generations seems roughly typical of these models (though note
that the 170 thousand configurations with no tilts have even
numbers of generations except in a few cases with \c-branes, so in the total set of configurations odd
generations are less frequent).  

Almost all (4704) of the 3-generation configurations have $SU(3)
\times SU(2)$ realized by 3 \b-branes of one kind and 2 \b-branes of
another kind.
An example of one of the 3\b $\times$ 2\b\ configurations realizing
$SU(3) \times SU(2)$ with 3 generations of matter in the bifundamental
of the gauge groups is
\begin{equation}
  3 \times (0, 9, 1, 0)_{(3, \widetilde{1/2}; 1, 3;0, -1)}
+ 2\times (0, 13, 0, 1)_{(1, \widetilde{1/2}; 0, -1; 1, 13)}
\label{eq:b32-3}
\end{equation}
This combination of branes requires at least one \a-brane to bring the
tadpoles down.  Over a dozen individual \a-branes
can be combined with (\ref{eq:b32-3}) to reduce all the tadpoles to
below the bound.  One example of such an \a-brane
is
\begin{equation}
(6, -45, 5, 6)_{(3, \widetilde{1/2}; 1, -3; 2, -5)}
 \label{eq:b32-a}
\end{equation}
Just including this \a-brane with the branes (\ref{eq:b32-3}) and
including 2 \c-branes with nonzero tadpoles $P$ as fillers gives a
model with total gauge group (before removing anomalous U(1) factors)
$U(3) \times U(2) \times U(1) \times Sp (2)$.  Solving the SUSY
equations (that is, solving (\ref{eq:SUSY-1}) for integer moduli with
the branes (\ref{eq:b32-3}, \ref{eq:b32-a}) using $\tilde{m}_1$ in place of $m_1$) shows that the
moduli for this model are $h = 4, j = 279, k = 26, l = 2$ (well
outside the range studied in \cite{Gmeiner}).  Typical of such
constructions, this model has additional exotic massless chiral matter
fields charged under the $SU(3) \times SU(2)$ part of the gauge group,
coming from nonzero intersections between the branes (\ref{eq:b32-3})
and the other branes in the model.  So this is not a realistic model
of nature, but provides an example of one of the many ways that branes
can consistently combine to form models with gauge group containing
$SU(3) \times SU(2)$ as a subgroup with 3 generations of ``quarks''.
All of the 4760 configurations with 3 generations are compatible with
at least one \a-brane configuration (and generally many) in such a way
as to satisfy the tadpole and SUSY constraints.  One might also worry
about K-theory constraints for complete models, but these may be
weaker when at least one torus is tilted \cite{Marchesano-thesis}.  We
will not say more here about these 4760 realizations of $SU(3) \times
SU(2)$ with 3 generations of quarks, leaving this to further work.
Clearly, however, these models form a good starting point for a
systematic analysis of models with features of the standard model.

We have focused here on the case of a single tilted torus, where
generally the $SU(3)$ and $SU(2)$ components of the gauge group are
both realized by \b-branes.
It is also worth considering the situation where the $SU(2)$ comes
from a \c-brane on the orientifold plane with $Sp(1)$ gauge group.  In
this case, the intersection number $I_{32}$ between the branes forming
the $SU(3)$ and the \c-brane can be odd even with no tilted tori.  We
have found that this intersection number is 3 in 49 distinct cases.
Of these 49, 4 have the $SU(3)$ realized by an \a-brane (in each case
having a tadpole $> 8$ requiring another \a-brane to fix tadpoles),
and 45 have the $SU(3)$ realized by a \b-brane.  One of the simplest
examples, where the \b-brane is $(0, 1, 9, 0)_{(1, 3; 3, 1; 0, -1)}$
and the \c-brane is $(0, 0, 0, 1)_{(0, 1; 0, -1; 1, 0)}$ was used in
\cite{cim, cl3, Marchesano-Shiu} to construct a semi-realistic
3-generation model.  Most of the other examples have much larger
tadpoles, such as realizing the $SU(3)$ by $3 \times (0, 2, 39, 0)_{(1,
13; 3, 2; 0, -1)}$, which obviously requires an additional \a-brane
with large negative $R$ tadpole.

A more complete analysis would be needed to determine which of the
realizations considered in this subsection can be completed to models
with additional branes saturating the tadpole conditions and the
K-theory constraints.  One could in principle look for further
structure reminiscent of the standard model in a straightforward
fashion by including the $U(1)$ and possibly extending the gauge group
further and computing the resulting possibilities for further matter
content.  We leave this endeavor to further work.

\subsection{Realizations of $SU(N) \times SU(2) \times SU(2)$}
\label{sec:n22}

As a further application of the method, we have constructed all
realizations of the gauge group $SU(N) \times SU(2) \times SU(2)$ for
various values of $N$.
This gives a picture of how adding additional components to the gauge
group increases the number of constructions including $SU(N)$.  Also,
the case $N = 4$ is relevant to construction of semi-realistic
Pati-Salam models as discussed in Subsection \ref{sec:models} .

The number of realizations of this group for various values of $N$ is
plotted in Figure 6.  For example, the number of distinct
realizations of $SU(4) \times SU(2) \times SU(2)$ is 587,704.
Starting with $N = 4$, the majority of solutions are of the form $NC
\times 2B \times 2B$ with a hidden \a-brane, for reasons similar to
those discussed above for the $SU(3) \times SU(2) \times U(1)$ models.
Starting from $N=8$, the $U(N)$ brane has to be a \c-brane (since a
stack of 8 or more \b-branes would oversaturate two of the tadpoles,
and one or more extra \a-branes can only compensate for one of the
excess tadpoles).  There is
thus a steady reduction in the number of configurations as $N$ increases.  The
hidden \a-brane with the most negative possible tadpole is of the form
$(P,Q,R,S) = (-512,8,8,8)$.  Thus at $N=518$, there is only one
realization, in which the group $SU(518) \times SU(2) \times SU(2)$
arises as a subgroup of $Sp(520)$ composed of 520 identical \c-branes
with tadpoles $(1,0,0,0)$.   There are no brane constructions
compatible with tadpole constraints for $N > 518$.

\begin{figure} \label{22nFig}
\begin{center}
\includegraphics[width=9cm]{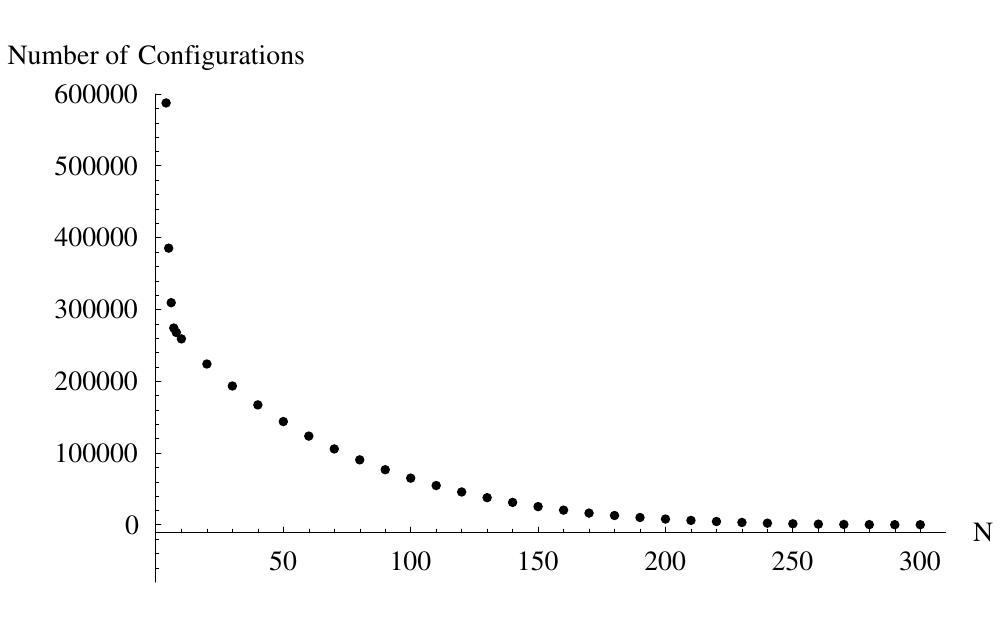}
\end{center}
\caption{Number of distinct
brane configurations giving $SU(N) \times SU(2) \times SU(2)$ (untilted tori).}
\end{figure}

\subsection{K-theory constraints}
\label{sec:K-theory}

Because we have focused here on constructing distinct brane
configurations of a given $G$ as a subgroup of the full gauge group,
and not on the details of how such configurations can be expanded to
complete models saturating the tadpole constraints, the K-theory
constraints do not apply directly to the configurations we have
constructed.  To test the K-theory constraints for a given
configuration, we must first add additional \b- and \c-branes to the
configuration to give a complete model which saturates the tadpoles,
and then test K-theory for this complete model.  For any given
configuration there may be many completions to a full model, and
consequently many
opportunities for satisfying the K-theory constraints.  We find,
however, that for some configurations there are no completions at all
to models satisfying the full K-theory constraints. Additionally, these
constraints have a stronger impact on the range of allowed
configurations further out in the tail of the distribution where the
dominant set of configurations arises from a single \a-brane with a very
negative tadpole.  This enhanced K-theory suppression of the tail
arises in part from the smaller number of completions possible from a
given realization of $G$ to a complete model saturating the tadpoles,
and in part from discrete effects associated with the integer
constraints on the problem. \footnote{Thanks to Robert Richter for
suggesting to us the possible role of integer effects in increasing
K-theory suppression in the tail, and for helpful discussions on this
question.}

To get a sense of how the K-theory constraints affect the
distribution of models containing a particular realization of $G$, we
have considered the set of $SU(3) \times SU(2)$ brane realizations
through a combination of a stack of 3 \b-branes and a stack of 2
\b-branes, with an arbitrary \a-brane combination included to provide
``phase space''.  As discussed above, there are some 164,000 distinct
3\b\ $\times$ 2\b\ configurations which can be realized in this way in
a fashion consistent with supersymmetry and the tadpole constraints.
While in general there are multiple ways in which each such
realization can be expanded to a complete model (before checking
K-theory) by saturating the tadpole constraints, one such way of
building a complete model is simply including ``filler'' \c-type
branes to saturate the tadpole.  Such \c-branes do not affect the
K-theory conditions.  We have checked the K-theory constraints for the
\c-brane completions of this set of models.  We find that in fact the
K-theory conditions substantially constrain the range of valid
constructions.  While the distribution still has a long tail, only
3636 of the 164,000 \c-brane  completions of the 3\b\ $\times$ 2\b\ combinations are 
compatible with the K-theory constraints.  Furthermore, the K-theory constraint suppresses
configurations with larger tadpoles more strongly.  Of the allowed
configurations, the average of the maximum total positive tadpole for
the \b-branes is around 32, and the maximum of this quantity is 68,
much smaller than for the distribution without K-theory constraints
shown in Figure 3.  It is interesting to note that K-theory reduces
the number of models possible with a single \a-brane.  Whereas before
the K-theory constraints are imposed all but 2 of the 164,000 \b-brane
3-2 combinations can be constructed with a single extra \a-brane, when
K-theory is imposed on the \c-brane completions of the configurations, only 3457
\junk{3523} of the 3636 distinct models are compatible
with a single \a-brane.   Including other
possible completions by  \b-brane combinations to saturate
the tadpole may increase the number of allowed configurations
substantially.  So the number of distinct ways in which $SU(3) \times
SU(2)$ can be realized by \b-branes which can be completed through
other branes to a full model satisfying SUSY, tadpole, and K-theory
constraints lies somewhere between the lower bound of 3636 (where only extra
\a-branes and \c-branes are added to give a model satisfying K-theory
constraints) and the upper bound of 164,057 (where arbitrary branes
can be added and K-theory constraints are not checked).  Nonetheless,
this analysis suggests that, while the greatest diversity of models is
still in the tail of the distribution when K-theory constraints are
included, this tail may be truncated substantially by the discrete
K-theory constraints when considering complete models of brane
combinations saturating the tadpole constraints.

The reason that the K-theory constraints substantially suppress
configurations containing a single \a-brane with a large negative
tadpole can be seen from the form of the winding numbers of such
branes.  All \a-branes with no tadpoles $Q, R, S > 8$, and one
negative tadpole $P < -112$ have the form (with a canonical choice of signs)
\begin{equation}
(P, Q, R, S)_{(n_1, m_1; n_2, m_2; n_3, m_3)} =
(-abc, a, b,c)_{(a, 1; b, 1; -c, -1)} \,.
\label{eq:brane-form}
\end{equation}
Such branes contribute a 1 to the first K-theory constraint in
(\ref{eq:K-theory}), from $m_1 m_2m_3 \equiv 1 \ ({\rm mod}\ 2)$.
A \b-brane can only contribute to this K-theory constraint if it has
all $m_i \neq 0$, so it has $P = 0$.  Such branes reduce the available
``phase space'' in winding numbers without contributing many
possibilities.  In particular, forming an $SU(3) \times SU(2)$ from
\b-branes where the 3 branes must all have $P = 0$ (necessary to fix
the first K-theory constraint in the absence of other \b-branes)
requires that the \a-brane have $Q, R, S$ tadpoles adding to at most
16 (we need $3 \times 2$ for the $SU(3)$, and another 2 for the
$SU(2)$).  The most negative tadpole $P$ at which this occurs is for
the brane of the form (\ref{eq:brane-form}) with tadpoles (-150, 5, 5,
6), and here supersymmetry rules out the \b-brane with tadpoles (0, 1,
1, 0).  All single \a-branes with $P < -112$ in combination with 3\b\ $\times$ 2\b\
and no further \b-branes are similarly ruled out, so we immediately
see that without at least one additional \b-brane, none of the
\b-brane combinations with maximum total tadpole above 120
contributing to Figure 3 can satisfy K-theory.  We defer a systematic
treatment of all models satisfying the K-theory constraints to further
work.  We note, however, that the brief discussion here demonstrates that the
K-theory constraints significantly decrease the number of
configurations associated with large tadpoles, although even the lower
bound given here for configurations satisfying K-theory still
represents a substantial range of models at moderate tadpoles of order
$30-50$.  We discuss further aspects of the reduction of the range of
allowed constructions in Section \ref{sec:comparison}.

\subsection{Comparison with previous results on IBM model-building}
\label{sec:models}

Many authors have constructed models with various features of the
standard model within the framework in which we are working of
intersecting branes on the $\orbifold$ orientifold.  
While the emphasis of this paper is on developing general tools which
can be useful either for model-building or for understanding the
distribution of models in this corner of the landscape, it is useful
to make contact with more detailed model-building results by checking
that the specific models found in earlier work are contained within
the larger classes of configurations constructed here.
We have checked
that various supersymmetric standard-model-like constructions in the
literature  which include $SU(3) \times SU(2)$ realizations with 3
generations of quarks are contained within our list of 4760 such
configurations.  For example, in \cite{csu1, csu2}, Cveti\v{c}, Shiu and
Uranga identified a model
containing $SU(3) \times SU(2)$
and 3 generations  of quarks on the orientifold with one tilted torus
using the configuration of \b-branes (in our
notation, with the contribution of $\tilde{m}_1$ to the tadpole doubled)
\begin{equation}
3 \times (1, 0, 1, 0)_{(1, \tilde{1/2}; 1, 0; 1, -1)}
+ 2 \times (1, 0, 0, 3)_{(1, \tilde{3/2}; 1, -1; 1, 0)}
\end{equation}
as well as additional branes completing the tadpole condition and
giving a ``semi-realistic'' spectrum including an additional gauge
sector and exotic chiral matter fields.  The intersection numbers
between these branes give $I_{32} = 1, I_{32'} = 2$ for a total of 3
generations of ``quarks''.  We have checked that this \b-brane
configuration is in our list of 4760 such configurations.  The authors
of \cite{csu1, csu2} restricted attention to brane configurations
composed completely of \b-branes, without any extra \a-branes in the
configuration.  As discussed above, this severely restricts the range
of possible models.  Precisely 10 of the 4760 realizations we
found of $SU(3) \times SU(2)$ with 3 generations of quarks can be
constructed without \a-branes somewhere in the configuration.  Note
that the model found in \cite{csu1, csu2} has additional
standard-model like features which may not be realizable by adding
branes to all the 4760 $SU(3) \times SU(2)$ brane configurations with
3 quark generations.  We leave a more detailed phenomenological
analysis of the range of complete models in which these 4760 brane
configurations can be embedded to further work.

Another popular approach to constructing semi-realistic IBM models
involves finding a brane configuration giving a Pati-Salam $SU(4)
\times SU(2)_L \times SU(2)_R$ group as part of the gauge group, with
various additional branes completing the tadpole conditions, and
giving additional gauge fields and exotic or hidden matter fields.  
In \cite{cim}, a model with 3 generations of matter fields charged
under the $SU(4)$ and each of the $SU(2)$'s was constructed using the
\b-\c-\c\ brane combination
\begin{equation}
4 \times (9, 1, 0, 0)_{(1, 0; 3, 1; 3, -1)}
+ 1 \times (0, 0, 1, 0)+ 1 \times (0, 0, 0, 1) \,.
\label{eq:ms-c}
\end{equation}
We have confirmed that this configuration appears in our list of over
$5\times 10^{5}$ realizations of $SU(4) \times SU(2) \times SU(2)$.  As pointed
out in \cite{Marchesano-Shiu}, this brane configuration can be
realized in a way compatible with SUSY and K-theory constraints in the
presence of the ``extra'' \a-branes
\begin{equation}
(-24, 2, 3, 4)_{(-2, -1; 3, 1; 4, 1)} +
(-24, 2, 4, 3)_{(-2, -1; 4, 1; 3, 1)} \,.
\label{eq:ms-a}
\end{equation}
(The addition of \a-branes to this model to fix the tadpoles was also discussed in \cite{cl3}.)

The use of \c-branes in (\ref{eq:ms-c}) makes possible the
construction of a 3-generation model without tilted tori.  This can be
realized in other ways; for example in \cite{Douglas-Taylor} a
Pati-Salam model of this type was found where the $SU(4)$ comes from
the \a-brane
\begin{equation}
4 \times (-3, 3, 1, 1)_{(3, 1; 1, 1; -1, -1)}
+ 1 \times (0, 0, 1, 0) + 1 \times (0, 0, 0, 1)
\label{eq:dt-422}
\end{equation}
where the extra \a-brane $(6, -4, 2, 3)_{(2, 1; 1, -1; 3, -2)}$ fixes
the tadpole excess, and further \b- and \c-branes can be added to
saturate the tadpole conditions while containing SUSY and satisfying
the K-theory constraints.
The 4-2-2 configuration in (\ref{eq:dt-422}) also appears in our
complete list of such constructions.

The Pati-Salam models just described use \c-branes and rectangular
tori to realize the $SU(2)$ parts of the model.  In \cite{cll}, Cveti\v{c}
Li and Liu performed a systematic search for Pati-Salam models where
the full $SU(4) \times SU(2) \times SU(2)$ gauge group arises from
$U(N)$'s on stacks of \a- and \b-branes, again with 3 generations of
matter in the ${\bf (4, 2, 1)}$ and ${\bf (\bar{4}, 1, 2)}$ representations
of the gauge group.  This was the first systematic search of this type
which included the possibilities of \a-branes and tilted tori.  They
found 11 models with the desired properties, including
additional constraints on the extra branes needed to saturate the
tadpole condition which assist with moduli stabilization and SUSY
breaking.  We have checked that the brane configurations giving
$SU(4) \times SU(2) \times SU(2)$ in
all 11 of these models appear in our
comprehensive list.

While a number of these ``semi-realistic'' intersecting brane model
constructions which have been analyzed in the literature share many
features of the observed standard model of particle physics, all
models constructed in this fashion so far also have some unrealistic
physical properties.  These models generally have exotic massless
chiral fermions charged under the standard model gauge group,
associated with non-vanishing intersection numbers between the branes
forming the standard model gauge group and extra branes which complete
the tadpole conditions.  Furthermore, while some progress has been
made towards incorporating fluxes to stabilize moduli (see for example
\cite{Marchesano-Shiu, cll2, Kumar-Wells, Marchesano-Shiu-2}), the intersecting brane
constructions have unstabilized moduli appearing as massless scalar
fields.  A more complete phenomenological model would need to
reproduce the standard model spectrum more precisely, as well as
stabilize moduli and give a complete picture of supersymmetry breaking
(a review of recent developments in using nonperturbative instanton
effects to resolve these issues is given in \cite{bckw}).  We also
know that the ingredients used in the type IIA orientifold
intersecting brane models cannot in principle give a realistic
cosmological scenario with many e-foldings of inflation, at least in the supergravity approximation without additional features like NS5-branes \cite{hktt}.
Thus, the models constructed in this fashion should be viewed at this
point as prototypes of string constructions of observable physics,
which have some desired features and may display interesting
characteristics common to more precisely tuned models.  The methods we
have developed in this paper can be used to isolate sets of IBM models
which have particular physical features, which may be useful in
further model-building studies.  These methods should also be
applicable in a wider range of compactifications, such as magnetized
brane constructions on a smooth Calabi-Yau where a much larger range
of constructions should be possible, and therefore more detailed
features of observed physics should be realizable.

\subsection{Other toroidal orbifolds}
\label{sec:other-orbifolds}

In this paper we are focused on the $\orbifold$ orientifold.  A number
of other toroidal orbifolds have been considered in the literature
\cite{IBM}.  The $\orbifold$ orbifold has a much richer structure than
many other toroidal orbifolds, in part because the moduli are not
completely fixed by the orbifold quotient.  In many other cases, the
quotient fixes the moduli and dramatically reduces the range of
possible models.  Nonetheless, other models have some interesting
features.  For example, the $T^6/\Z_4 \times\Z_2$ orientifold has
solutions with standard model-like chiral matter \cite{Honecker-z4}
and been used to study $SU(5)$ GUT models \cite{Cvetic-Langacker}.
Although in principle the methods of this paper could be used to study
the range of models available in other orbifold constructions, in many
cases the analysis is much simpler due to the absence of free moduli,
and there are generally fewer distinct constructions of physical
properties of interest.  Two potentially interesting models for study
are the $T^6/\Z_6$ and $T^6/\Z_6'$ toroidal orientifold models, which
have been shown to have phenomenologically interesting solutions
\cite{z6, z6p}.  A statistical analysis of solutions on these
orbifolds has also been carried out in \cite{z6-statistics, Gmeiner-Honecker}.
For the   $T^6/\Z_6'$ orientifold
Gmeiner and Honecker carried out a
complete analysis of solutions in
\cite{Gmeiner-Honecker}.  They found a large total number of models,
of order $10^{23}$.  The
exponentially large number of models in this case comes from the large
combinatorial number of ways in which a relatively small number of
distinct branes can be combined to saturate the tadpole conditions,
along with an exponential enhancement from exceptional cycles beyond
those on the bulk torus.  We describe in the next section how a
similar feature to the first of these effects affects the distribution
of complete models for the $\orbifold$ orientifold we are studying
here at small integer moduli.  It seems that in the $T^6/\Z_6'$ model,
while there are many total models, the number of distinct
constructions of a particular structure such as the standard model
gauge group $G_{123}= SU(3) \times SU(2) \times U(1)$ is  smaller
than in the $\orbifold$ case, even though the number of ways in which
``extra'' branes can be added to realize a complete model containing
any given realization of $G_{123}$ is large.  It would be interesting
to use some of the methods developed here for a further study of
these and other
orbifold constructions.
To understand where the
diversity in constructions can be found for the $\orbifold$
orientifold, it is helpful to compare our results to the analysis of
Gmeiner {\it et al.} in \cite{Gmeiner}, to which we now turn.

\section{Diversity in the ``tail'' of the IBM distribution}
\label{sec:comparison}

In \cite{Gmeiner}, Gmeiner, Blumenhagen, Honecker, L\"ust, and Weigand
undertook an ambitious computational effort to scan over all complete
solutions to the tadpole, SUSY, and K-theory constraints.  
Their
approach was to scan over the integer moduli 4-vector $\vec{U} =(h, j,
k, l)$.  For each set of moduli, they computed the complete set of
possible brane combinations compatible with all constraints.
Their analysis proceeded up to moduli with
norm $| U | = 12$.  Their algorithm became exponentially difficult as
$| U |$ increased, but the number of solutions seemed to be decreasing
for larger $| U |$, so the numerical evidence indicated that they had
scanned the majority of all possible solutions, with a small fraction
remaining in the tail of the distribution at larger $| U |$.

\begin{figure}[h] \label{fig:modDist}
\begin{center}$
\begin{array}{cc}
\includegraphics[width=10cm]{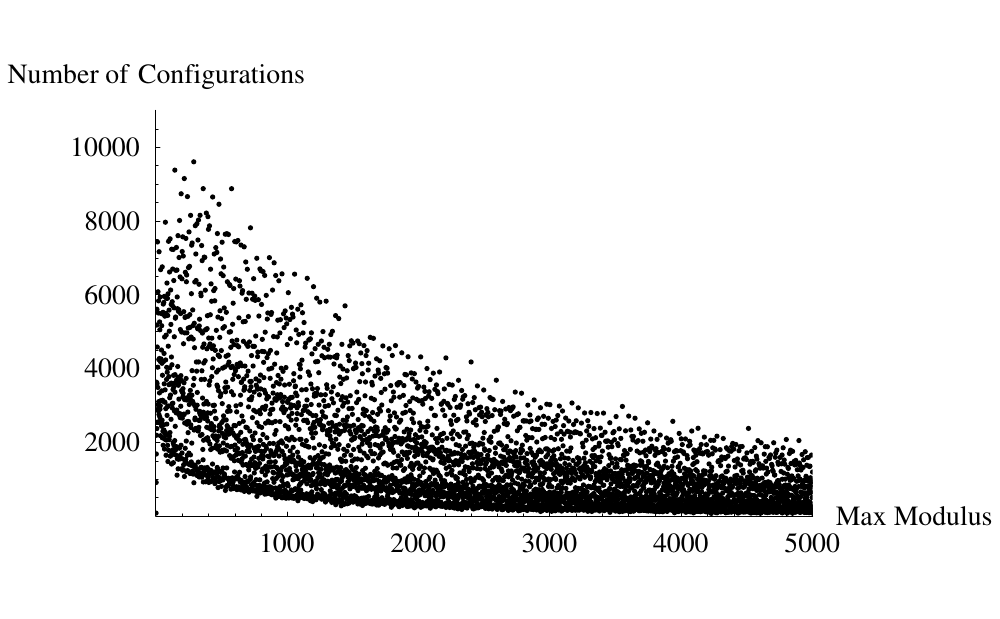} \\
\includegraphics[width=10cm]{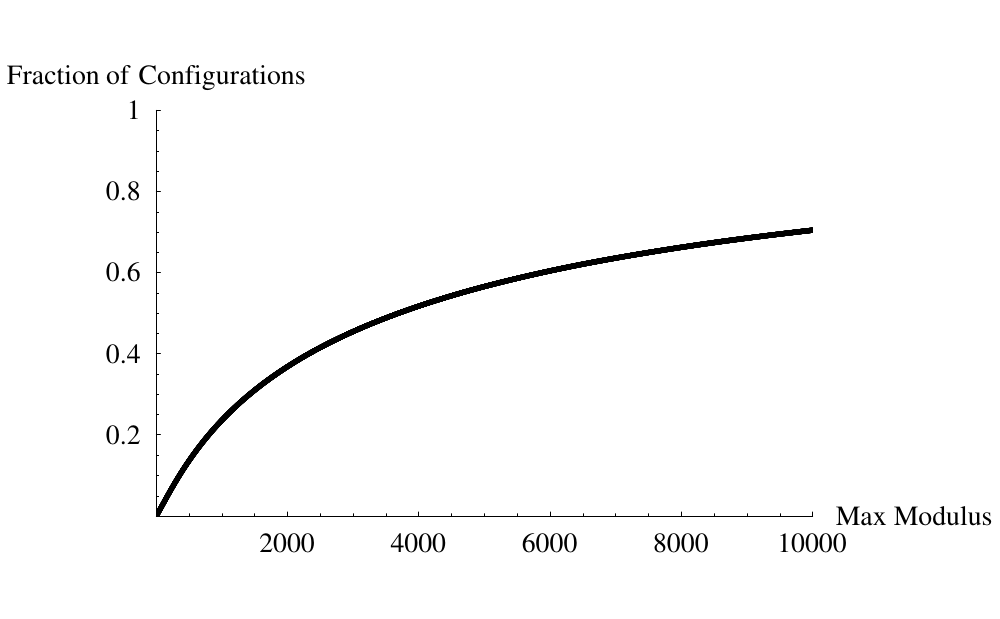}
\end{array}$
\end{center}
\caption{The upper plot shows the distribution of the number of brane
  configurations realizing gauge group $G_{321}$ as a function of the maximum modulus.  On the lower
plot the $x$-axis is the maximum modulus and the $y$-axis is the
fraction of configurations with $G_{321}$ that have all moduli smaller than $x$.  }
\end{figure}

At first sight, the results of this numerical analysis seem rather at
odds with the conclusions we have reached in this paper, which are
that the vast majority of distinct realizations of specific gauge
groups like $G_{321} = SU(3) \times SU(2)\times U(1)$ occur in
combination with \a-branes with large negative tadpoles, which are
associated with large integer moduli.  To verify that most of the
configurations we have found occur at large moduli, we plot in Figure 7
the distribution of the maximum integer modulus (max($h,j,k,l$) $\leq | U|$) of a subset of the
configurations we found that contain $G_{321}$.  The subset for which
the moduli are computed and plotted consists of those configurations
in which the moduli are uniquely determined either from the
realization of $G_{321}$, or by the branes realizing $G_{321}$ in
combination with all possible \a-brane combinations which lead to
undersaturation of the tadpole conditions.  In the latter case, we use
the minimum across all compatible extra \a-brane combinations of the maximum
modulus.  This subset with fixed moduli represents 97\% of the
13.7 million distinct realizations of $G_{321}$.  The remaining brane
configurations realizing $G_{321}$, not included in the plot, either
do not fix the moduli at all, or
require additional branes to fix the moduli for some compatible
\a-brane combinations.

The results shown in the plots in Figure 7 give
definitive confirmation of the story described above.  Most
realizations of $G_{321}$ arise from brane configurations which fix
the moduli in the range $500 < | U | < 20,000$.  Less than one half of
1\% of the distinct realizations of $G_{321}$ which fix the moduli
appear in the range $| U | \leq 12$ scanned in \cite{Gmeiner}.

How then can these results be compatible?  
To understand the resolution of this apparent discrepancy it is helpful to
consider the distinct nature of the questions asked in performing
these two analyses.  In Gmeiner {\it et al.}'s work, they were
scanning over all possible brane configurations which completely
saturate the tadpole constraints.  Thus, at any particular value of
the moduli they include all models, which for small moduli can contain
a very large number of combinations of a relatively small number of
distinct branes.  Even if the number of combinatorial possibilities is
large at small moduli, the number of distinct realizations of any
particular gauge subgroup may be relatively small.  On the other hand,
in our analysis we are simply looking for distinct realizations of a
gauge subgroup $G$, not necessarily counting the number of ways in
which extra branes can be added to the branes forming $G$ to form a
complete model.  The number of distinct realizations of $G$ at small
moduli can be relatively small, while there can be an exponentially
large number of ways of completing $G$ to a complete model.
Furthermore, for more complex groups like $G_{321}$ there may be fewer
moduli at which such groups can be realized by any brane combination.
Thus, the apparent discrepancy between our results and those of
\cite{Gmeiner} can be understood from the realization that while the
number of complete models at fixed moduli decreases fairly rapidly for
small moduli, the tail of this distribution is extremely long.  While
the tail of the distribution may have fewer total models than the
``bulk'' at small moduli, it can contain a much wider diversity of
realizations of any fixed gauge group $G$.

To check this explanation, we have considered several examples of
small moduli and large moduli, and determined the total numbers of
models consistent with those moduli and the tadpole and SUSY
constraints.  We found that indeed for small moduli, there can be many
distinct solutions to the constraints.  For example, for the integer
moduli $(h,j,k,l) = (1,2,3,4)$ there are 39,871 models coming from
brane configurations which satisfy SUSY constraints and saturate the
tadpole conditions (without considering K-theory constraints).  For
such small moduli, however, generally there are few, if any, distinct
realizations of any particular group $G$, such as $G_{321}$.  Those
realizations of $G$ which appear at small moduli may be consistent
with many possible completions to total models through addition of
different types of extra branes, but these represent a small fraction
of the total number of realizations of $G$.  At large moduli, on the
other hand, generically there are few if any consistent models, with
or without any particular gauge subgroup $G$.  For example, for the
moduli $(h,j,k,l) = (2, 26, 91, 1129)$ there are only 29 models in
total (without considering K-theory constraints), but one of these
does contain $G_{321}$.  A clear example of the dichotomy between
total number of models and numbers of distinct realizations of $G$ is
given by the simplest case, $G = U(1)$.  At moduli $(h,j,k,l) =
(1,2,3,4)$, there are 46 distinct branes, associated with 46 distinct
realizations of $U(1)$, which combine to form the
39,871 total models found there, while at moduli $(h,j,k,l) = (2, 26,
91, 1129)$, there are 14 distinct branes forming the 29 total models.
Thus, while the number of total models is dramatically reduced at
higher moduli, the number of distinct brane configurations realizing
$G$ does not decrease at the same rate.

Finally, as discussed in \ref{sec:K-theory}, K-theory constraints
reduce the number of possible models in the tail.  In \cite{Gmeiner} it
was found that K-theory constraints generically reduced the number of
allowed models by a factor of something like 5.  At the moduli $(1, 2,
3, 4)$ discussed above, 6004 of the roughly 40,000 total models are
compatible with the K-theory constraints, giving a reduction factor of
around 6.  At higher moduli, however, the suppression is stronger, as
suggested by the discussion in Section~\ref{sec:K-theory}.  While, for
example, at the moduli (2, 26, 91, 1129) discussed above, 14 of the 29
models are compatible with K-theory (including one containing
$G_{321}$), at other moduli containing models satisfying the tadpole
and SUSY constraints there are no models at all compatible with the
K-theory constraints.  For example, at moduli (35, 63,1, 3455) there
are 37 total models compatible with SUSY and tadpole cancellation, but
none of these models satisfy the K-theory constraints.  Again, the
tail of the distribution is suppressed in part by the K-theory
constraints, although it still seems to contain most of the diversity
of models.  To see how the K-theory suppression of the tail affects
the distribution of configurations in the moduli space, we have
checked the K-theory constraints for the \c-brane completions of the $SU(3) \times SU(2) \times
U(1)$ configurations graphed in Figure 7.  Only {\rm 385,889} of these
13 million models satisfy K-theory,
representing a reduction by a factor of more than 35.  Of those models
satisfying the K-theory constraints, roughly 90\% fix all moduli.  The distribution on the space of moduli is
graphed in Figure 8.  As in the discussion of Section
\ref{sec:K-theory}, this represents a lower bound on the range of
allowed models, as some of the configurations which do not satisfy the
K-theory constraints with a pure \c-brane completion
may be compatible with
K-theory constraints if \b-branes are used in the completion to models with saturated tadpoles.
Nonetheless, we expect that Figure 8 presents a more accurate picture
of the distribution on moduli space than Figure 7.  We see in Figure 8
that the K-theory constraints reduce the range of brane configurations
more strongly in the far end of the tail of the distribution.
Nonetheless, the tail is still quite large, with many models having
moduli of order 100-1000. Roughly 50\% have a maximum modulus greater
than 180. After imposing K-theory constraints on these models, some 5\% of $SU(3) \times SU(2) \times U(1)$
models lie within the region of moduli space scanned in
\cite{Gmeiner}.

\begin{figure}[h] \label{fig:modDist1}
\begin{center}
\includegraphics[width=10cm]{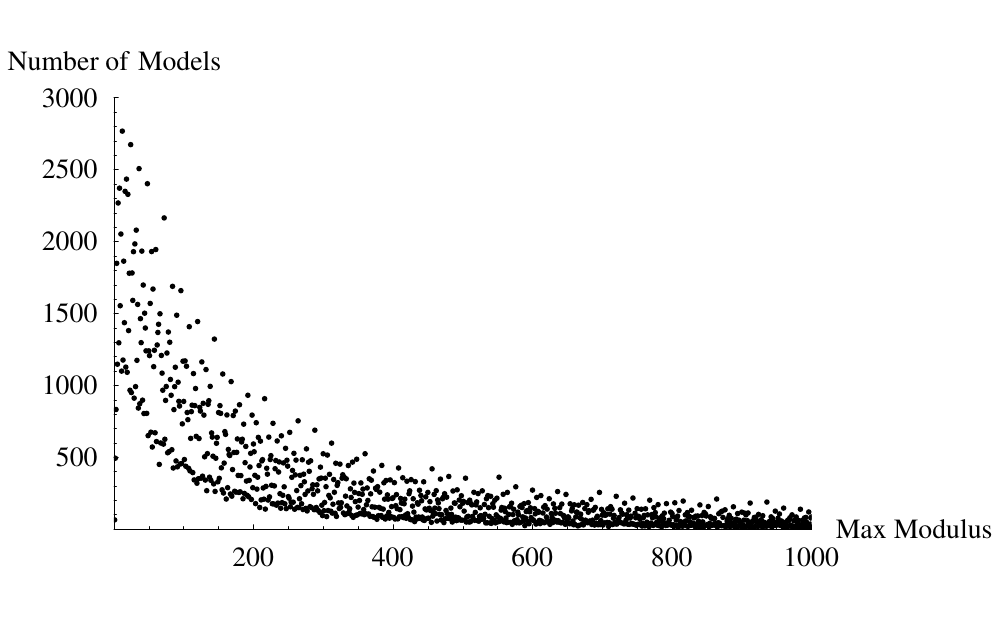} 
\end{center}
\caption{Distribution of a subset of models satisfying K-theory constraints. These models are formed from minimal \c-brane completion of the configurations realizing $G_{321}$ graphed in Figure 7.}
\end{figure}

Thus, while at first glance it seems that there is some apparent
contradiction between the results of our analysis and those of Gmeiner
{\it et al.} in \cite{Gmeiner}, in fact these results are completely
compatible.  The reconciliation of these different analyses lies in
the recognition that the distribution of models has a very long tail and there are fewer completions of a given realization of a gauge group
$G$ to a complete model in the tail. Therefore, even though the end
of the tail is somewhat suppressed by the K-theory
constraints, the greatest diversity of distinct realizations of
specific gauge groups or matter content lies in the tail of the
distribution, at large moduli.

Other large regions of the
landscape which have been studied from a statistical point of view
include Gepner models \cite{Gepner} and heterotic constructions
\cite{Dienes}.  It would be nice to have some general lessons which
are applicable to  diverse sets of vacua, and
the lessons learned here may have interesting ramifications for study
of other patches in the landscape.  From a model-building
point of view, it is clearly important to understand what structure is
needed to realize the greatest diversity of possible low-energy
theories.  From a general landscape point of view, it is important to
understand how to characterize the range of low-energy physics which
can be realized in a given string construction.  And from the point of
view of extracting general lessons from string compactifications
relevant for heuristic predictions about physics beyond
the standard model, the notion that the widest range of low-energy
physics theories may arise in regions where there are fewer possible
extensions to observable physics may have some bearing on our thinking
about how string theory relates to physics which will hopefully be
observed beyond the standard model.  For example, it may be that in
the tail of the landscape distribution, extra massive U(1) factors are
not quite as ubiquitous as in the bulk of the distribution.  Depending
on how finely tuned our physics must be, this might decrease our
expectation of seeing massive Z''s at the LHC.  Or it might not.  It
would be rather premature to take seriously any such speculation based
on our current extremely limited understanding of the nature of the
full string landscape.

\section{Conclusions}
\label{sec:conclusions}

In this paper we have developed a systematic approach to constructing
all intersecting brane models on a particular toroidal orientifold.
The key technical result is the determination of all allowed
combinations of branes with negative tadpole contributions, using
bounds on winding numbers of these branes arising from the tadpole
constraints and SUSY conditions.  Given these combinations of
``\a-branes'', the construction of models with any desired specific
features amounts computationally to a straightforward combinatorial
problem of polynomial complexity, since all other branes besides
the \a-branes must have positive tadpole contributions, and all
tadpoles have a fixed bound.

The methods developed here should generalize to other classes of
models.  For some models, such as other toroidal orientifolds
\cite{IBM} including the $\Z_6$ and $\Z_6'$ orbifold models studied in
\cite{z6-statistics, Gmeiner-Honecker}, and magnetized brane models on
K3 \cite{Kumar-Taylor}, there are no branes with negative tadpole
contributions and/or fewer free moduli in the problem, so the problem
of classifying solutions is simpler.  For more general models,
however, such as magnetized brane models on general Calabi-Yau
manifolds, we expect an analogue of \a-branes \cite{Douglas-Taylor},
with some negative tadpole contributions and a large number of moduli.
For such models, the methods developed here may prove useful in
gaining mathematical and computational control over the range of
low-energy theories accessible through various brane constructions.
Furthermore, the form of the mathematical problem addressed in this
paper is very similar to other classes of compactification problems,
such as flux compactifications, where a total tadpole constraint and
SUSY conditions must be solved to determine the range of allowed
constructions.  It may be that some general methods similar to those
developed here
may be useful in
addressing all of these types of vacuum classification problems.

More concretely, the analysis of this paper has given us a clear
picture of the overall structure of the space of supersymmetric
intersecting brane models on this particular toroidal orientifold.
The computation of all combinations of branes with negative tadpoles
makes possible a straightforward enumeration of all brane
configurations compatible with SUSY which realize any desired gauge
group and/or matter content.  We have explicitly enumerated ways in
which the gauge groups $U(N), SU(N) \times SU(2) \times SU(2), SU(3)
\times SU(2)$, and $SU(3) \times SU(2) \times U(1)$ can be realized.
We found that the great majority of these gauge group realizations are
associated with one or more \a-branes with large negative tadpoles,
which enables large winding numbers for the branes composing the gauge
group in question, and consequent large integer moduli.  The discrete
K-theory constraints suppress configurations with single \a-branes and
extremely negative tadpoles.  

We have found that the greatest
diversity in realizations of any given gauge group and matter
structure occurs in the ``tail'' of the distribution of models.  This
tail is characterized by large integer moduli, generally outside the
range encompassed by the systematic scan of \cite{Gmeiner}, and by
one, or sometimes several D-branes of type \a\ with a very negative
tadpole.  In this tail, at fixed moduli, the number of ways in which a
given realization of a small gauge group can be completed to form the
total gauge group is much smaller than in the ``bulk'' of the
distribution at small moduli, where there are many possible
configurations of extra brane sectors, which could be hidden or, more
usually, include additional chiral matter fields charged under the
desired gauge group.  We have found evidence for further suppression
of the tail from the K-theory constraints.  Nonetheless, the vast
majority of distinct realizations of any given small gauge group occur
in the tail of the distribution, at large moduli.  It is difficult to
make concrete statements about the larger string landscape based on
this one sample, but it seems plausible that this feature of
``diversity in the tail'' of the distribution may be valid in more
general classes of compactifications.  If so, this might give some
insight both into efficient strategies for realistic model-building,
and into the expected range of additional gauge group and matter
content (such as the number of Z''s) which may arise naturally from
string theory in generic extensions of the standard model.  It would
be interesting to understand the phenomenon discovered here of
diversity in the tail of the landscape distribution in terms of some
more general measure on the moduli space, along the lines of
\cite{Ashok-Douglas}.  Understanding more generally where the
diversity of low-energy physics arises in the distribution of models
on the landscape may lead to a useful refinement of the statistical
approach to the string landscape pioneered by Douglas in
\cite{Douglas}.

Another general lesson of the results in this paper is that what
appear to be ``typical'' features of a model depend strongly on the
prior assumptions made about the structure of the model.  If several
assumptions are made, each associated with a cut on the data, then
depending on the form of these assumptions, conclusions may depend
upon the order in which the cuts are made.  For example, if one first
makes the assumption that one is interested in a ``typical'' model in
the space of all consistent intersecting brane models on the
$\orbifold$ orientifold, and therefore restricts to models with small
moduli, and second makes the assumption that the gauge group contains
$G_{321} =SU(3) \times SU(2) \times U(1)$, then one reaches a much
more restrictive set of possible models then if one first makes the
assumption that the gauge group contains $G_{321}$, and then looks for
typical features among the 16 million or so distinct realizations of
$G_{321}$.  In particular, as the analysis of this paper shows, the
order of cuts in this case impacts the number of possible ``hidden
sectors'' which can complete the model with gauge group $G_{321}$ to
form a complete and consistent intersecting brane model satisfying the
tadpole and supersymmetry constraints.  These results suggest that
model-building by ``trial and error'' may be less productive in
generating models with interesting phenomenology than more systematic
searches which hone in on subsets of models with desired features
which may be found in atypical places in the parameter space of
models.  This lesson may be useful to keep in mind in other contexts
involving analysis of regions of the string landscape.

\section*{Acknowledgements}

It is a pleasure to thank Allan Adams, Ralph Blumenhagen, Mirjam
Cveti\v{c}, Michael Douglas, Florian Gmeiner, Gabriele Honecker, Vijay
Kumar, Dieter L\"ust, John McGreevy, Robert Richter, Gary Shiu, and
Timo Weigand for helpful discussions and comments.  Particular thanks
to Michael Douglas for collaboration on the work reported in
\cite{Douglas-Taylor} and further discussions which helped initiate
this work.  Particular thanks also to Gabriele Honecker and Robert
Richter for detailed comments on a preliminary version of this
manuscript.  Thanks also to the KITP for support and hospitality while
part of this work was being completed.  This research was supported by
the DOE under contract \#DE-FC02-94ER40818.  This research was also
supported in part by the National Science Foundation under Grant No.\
PHY05-51164.

\appendix
\appendix

\section{Appendix}
In this Appendix we show there are no \Ab-brane
configurations with 4 different negative tadpoles ($[p,q,r,s]$) when the tadpole constraint has $T=8$. We use the notation introduced in Section \ref{sec:a} and recall that the branes can be arranged so that the $p$ branes with the negative $P$ tadpole are first in the configuration, followed by the $q$ branes with a negative $Q$ tadpole, then $r$ branes with a negative $R$, and finally $s$ branes with a negative $S$. Additionally, within these groupings the negative tadpoles are ordered in increasing order of their absolute value. We write all tadpbole numbers as the absolute value of the tadpole contribution and explicity insert the minus sign. So, for instance we have that  $P_1 \leq P_2 \leq ...  \leq P_p$ and  similarly for $Q$, $R$, and $S $ tadpoles. Furthermore, for convenience of notation, we let $a=p, \  b=p+q,  \ c=p+q+r, \ d= p+q+r+s$ and 
without loss of generality, $p\geq q \geq r \geq s$. 

Let us look at branes
$b,c$, and $d$.  Since there are $p+r+s-2$ branes other than branes $c$ and $d$ with a tadpole other than
$Q$ that is negative, and since $Q_b$ is the largest of the negative
$Q$ tadpoles, the tadpole constraint gives
\begin{equation}
Q_{c}+ Q_{d} \leq 8 + q Q_b - (p+r+s-2).
\end{equation}

Using the identity $\frac{1}{x}+\frac{1}{y}\geq \frac{4}{x+y}$ we have that 
\begin{equation} \label{eq:Qcd}
\Ov{Q_c}+\Ov{Q_d} \geq \frac{4}{10+ q Q_b - (p+r+s)}.
\end{equation}

Looking at branes $a,b,c,d$ and using the tadpole constraint
\begin{equation}
Q_a + Q_c + Q_d \leq 8 + q Q_b - (p+r+s-3)
\end{equation}
we similarly get 
\begin{equation} \label{eq:Qacd}
\Ov{Q_a}+ \Ov{Q_c}+\Ov{Q_d} \geq \frac{9}{11+ q Q_b - (p+r+s)}.
\end{equation}

The analogous relation for the $P$'s gives
\begin{equation} \label{eq:Pbcd}
\Ov{P_b} + \Ov{P_c} + \Ov{P_d} \geq \frac{9}{11+ p P_a - (q+r+s)}.
\end{equation}

Recalling \ref{eq:pq8} which states
\begin{equation}
p+q \leq 7,
\label{eq:pq7}
\end{equation}
combined with the assumption $p\geq q \geq r \geq s$, gives 
\begin{equation} \label{eq:prs10}
p+r+s \leq 10.
\end{equation}

We are now going to show there are no solutions for the following three cases:\newline
a) $p \leq q +r + s -2$  \newline
b) $q \leq p+r+s - 6$ \newline 
c) $q=r=s=1$ \newline

a) Using $P_a \geq 1$ combined with the assumption $p \leq q +r + s -2$ we get 
$(9-p)P_a \geq 11 -(q+r+s)$.  Rearranging gives $9 P_a \geq 11 - (q+r+s) + p P_a$.  Using (\ref{eq:prs10}) along with $p \geq q$ shows us that $11- (q+r+s)>0$. Therefore the right side of (\ref{eq:Pbcd}) is greater than or equal to $1/P_a$, 
\begin{equation} \label{eq:9frac}
\frac{9}{11+ p P_a - (q+r+s)} \geq \frac{9}{9 P_a} = \frac{1}{P_a}.
\end{equation}

So using (\ref{eq:Pbcd}) we see that the sum of the inverses of the $Ps$ for branes $a, b, c$ and $d$ is greater than or equal to 0
\begin{equation} \label{eq:Ovfrac}
-\Ov{P_a} + \Ov{P_b} + \Ov{P_c} + \Ov{P_d} \geq 0.
\end{equation}

Analogous arguments show that the sum of the inverses of the $Q's$ for
these 4 branes is greater than or equal to 0, as is the sum of the
inverses of the $R's$ and $S's$.  If we now add the SUSY equations
(\ref{eq:SUSY-1b}) for branes $a, b, c$ and $d$, the right side is $0$
while the left side is greater than or equal to $0$. For there to be
equality, (\ref{eq:Ovfrac}) and (\ref{eq:9frac}) must be equalities,
requiring $(9-p)P_a = 11- (q+r+s)$. This means $P_a = 1$ and
$p+2=q+r+s$, implying (using (\ref{eq:pq7})) that $r=s=1$. The analogous equations to
(\ref{eq:Ovfrac}) for the $Q's, R's$ and $S's$ must also be equalities
and thus $P_a = Q_b = R_c=S_d=1$ and $p=q=r=s=1$. The SUSY and tadpole
constraints prevent this from occuring. Hence, there are no solutions
in case (a).

b) Using $Q_b \geq 1$ combined with (\ref{eq:prs10}) we get for the right side of (\ref{eq:Qcd})
\begin{equation}
\frac{4}{10+ q Q_b - (p+r+s)} \geq \frac{4}{10 +q - (p+r+s)} \Ov{Q_b}.
\end{equation}

By assumption $10 +q - (p+r+s) \leq 4$. Thus with use of (\ref{eq:Qcd}) we get 
\begin{equation}
-\Ov{Q_b} +\Ov{Q_c} + \Ov{Q_d} \geq 0.
\end{equation}

Similarly the sum of the inverses of the $R's$ and $S's$ from branes
$b, c$, and $d$ are greater than or equal to $0$.  Adding the SUSY
equations (\ref{eq:SUSY-1b}) from branes $b, c$ and $d$ we see there
are no solutions in case (b).  

c) We begin with the case of $[2,1,1,1]$.  Using (\ref{eq:Qacd}) we
see that $\Ov{Q_2}-\Ov{Q_3}+\Ov{Q_4}+\Ov{Q_5} >0$.  Similarly the sum
of the inverses of the $R's$ and $S's$ for the last 4 branes is
greater than 0.  Upon adding the SUSY equations for these four branes
we see that we must have $-\Ov{P_2}+\Ov{P_3}+\Ov{P_4}+\Ov{P_5} <
0$. Applying (\ref{eq:Pbcd}) this means $\frac{9}{11+2P_2 -3} <
\frac{1}{P_2}$.  Rearranging shows $P_2 <8/7$. Combined with the
assumption that $P_1\leq P_2$ gives $P_1=P_2=1$.  From (\ref{eq:Qcd})
we get $\Ov{Q_4} + \Ov{Q_5} \geq \frac{4}{7}\Ov{Q_3}$.  Similarly,
$\Ov{S_3}+\Ov{S_4} \geq \frac{4}{7}\Ov{S_5}$ and $\Ov{R_3} + \Ov{R_5}
\geq \frac{4}{7}\Ov{R_4}$.  Now adding the SUSY equations for the last
three branes we find that
\begin{equation} \label{eq:equal1}
\frac{1}{P_3}+\frac{1}{P_4}+\frac{1}{P_5} \leq \frac{3}{7}(\frac{j}{Q_3}+\frac{k}{R_4}+\frac{l}{S_5}).
\end{equation}
Using similar arguments to those used to derive (\ref{eq:Qacd}) 
(and the discrete nature of the integers)
we get 
\[
\Ov{Q_1}+\Ov{Q_2} + \Ov{Q_4}+ \Ov{Q_5} \geq \frac{11}{6}\Ov{Q_3}.
\]
Note that equality for this equation occurs at, for example, $Q_1=3$
and $Q_2 = Q_4= Q_5 = 2$. The analogous equations hold for the $R's$
and $S's$.  Now we add the SUSY equations for all 5 branes and
remember that $P_{1}=P_{2}=1$.  This gives
\begin{equation} \label{eq:equal2}
2- (\frac{1}{P_3}+\frac{1}{P_4}+\frac{1}{P_5})\geq \frac{5}{6}(\frac{j}{Q_3}+\frac{k}{R_4}+\frac{l}{S_5})
\end{equation}
Combining the two inequalities (\ref{eq:equal1}) and (\ref{eq:equal2}), we find that 
$\frac{1}{P_3}+\frac{1}{P_4}+\frac{1}{P_5} \leq \frac{36}{53}$.  Subject to the tadpole condition $P_3+P_4+P_5\leq 10$, this has no solutions. 

For $p=3$ and $p=4$ we can use an analogous argument to the one above.  For $p\geq 5$ we notice that $ -\frac{1}{Q_{a+1}} +\frac{1}{Q_{a+2}} +\frac{1}{Q_{a+3}} >0$ and similarly for the $Rs$ and $Ss$.  Adding the SUSY equations for the last 3 branes, we see that there can be no solutions. 
\newline

There are only 5 remaining cases not included in $a -c$. They are $[3,2,1,1]$, $[4,2,1,1]$, $[5,2,1,1]$, $[4,3,1,1]$, and $[4,2,2,1]$. Using (\ref{eq:Qacd}) and (\ref{eq:Pbcd}) in analogous way as before,  for all of them we find $P_a=1$. Use of FWNB (\ref{eq:FWNB}) and TCSB (\ref{eq:TCSB}) imposes tight constraints on the other tadpoles, allowing for a systematic search which reveals there are no solutions in any of these cases.

\end{document}